%% file: paper.tex
\documentclass[number,sort&compress,twocolumn,5p]{elsarticle}
\usepackage[binary-units=true,separate-uncertainty=true]{siunitx}
\usepackage{lineno,hyperref}
\usepackage{tikz}
\usepackage{amssymb,amsmath,mathtools}
\usepackage{caption}
\usepackage{xspace}
\usepackage{booktabs}
\usepackage{siunitx}
\usepackage{textcomp}
\usepackage{booktabs}
\usepackage{subcaption}
\usepackage[percent]{overpic}
\usepackage[list=true,listformat=simple,margin=10pt]{subcaption}
\usepackage{hepnames}
\usepackage{lineno}
\usepackage{placeins}

\journal{Nucl. Instr. Meth. A}

\begin{document}
\begin{frontmatter}
  \title{Test-beam characterisation of the CLICTD technology demonstrator - a small collection electrode High-Resistivity CMOS pixel sensor with simultaneous time and energy measurement}

\author[cern]{R.~Ballabriga}
\author[cern]{E.~Buschmann}
\author[cern]{M.~Campbell}
\author[cern]{D.~Dannheim}
\author[cern]{K.~Dort\corref{corr}\fnref{ugiessen}}
\ead{katharina.dort@cern.ch}
\author[cern]{N. Egidos\fnref{barca}}
\author[desy]{L. Huth}
\author[cern]{I.~Kremastiotis}
\author[cern]{J.~Kr\"oger\fnref{heidelberg}}
\author[cern]{L.~Linssen}
\author[cern]{X.~Llopart}
\author[cern]{M.~Munker}
\author[cern]{A.~N\"urnberg\fnref{kit}}
\author[cern]{W.~Snoeys}
\author[desy]{S.~Spannagel}
\author[cern]{T.~Vanat\fnref{nowDesy}}
\author[cern]{M.~Vicente}
\author[cern]{M.~Williams\fnref{glasgow}}

\address[cern]{CERN, Geneva, Switzerland}
\address[desy]{DESY, Hamburg, Germany}

\cortext[corr]{Corresponding author}

\fntext[ugiessen]{Also at University of Giessen, Germany}
\fntext[barca]{Also at University of Barcelona, Spain}
\fntext[heidelberg]{Also at University of Heidelberg, Germany}
\fntext[kit]{Now at KIT, Karlsruhe, Germany}
\fntext[nowDesy]{Now at DESY, Hamburg, Germany}
\fntext[glasgow]{Also at University of Glasgow, U.K.}

\begin{abstract}

The CLIC Tracker Detector (CLICTD) is a monolithic pixel sensor.
It is fabricated in a 180\,nm CMOS imaging process, modified with an additional deep low-dose n-type implant to obtain full lateral depletion. 
The sensor features a small collection diode, which is essential for achieving a low input capacitance. 
The CLICTD sensor was designed as a technology demonstrator in the context of the tracking detector studies for the Compact Linear Collider (CLIC).
Its design characteristics are of broad interest beyond CLIC, for HL-LHC tracking detector upgrades.
It is produced in two different pixel flavours: one with a continuous deep n-type implant, and one with a segmented n-type implant to ensure fast charge collection.
The pixel matrix consists of $16\times128$ detection channels measuring \SI{300 x 30}{\micro m}. 
Each detection channel is segmented into eight sub-pixels  to reduce the amount of digital circuity while maintaining a small collection electrode pitch. 
This paper presents the characterisation results of the CLICTD sensor in a particle beam.
The different pixel flavours are compared in detail by using the simultaneous time-over-threshold and time-of-arrival measurement functionalities. 
Most notably, a time resolution down to $(5.8 \pm 0.1)$\,ns and a spatial resolution down to $(4.6 \pm 0.2)$\,\SI{}{\micro m} are measured.
The hit detection efficiency is found to be well above \SI{99.7}{\percent} for thresholds of the order of several hundred electrons. 

\end{abstract}


\end{frontmatter}


\section{Introduction}
\label{sec:introduction}
\input{introduction}


\section{The CLICTD monolithic sensor}
\label{sec:clictd}
 \input{clictd}

\section{Samples and readout}
\label{sec:assemblies}
\input{assemblies}

\section{Laboratory characterisation}
\label{sec:configurations}
\input{configurations}


\section{Test-beam and analysis setup}
\label{sec:testbeam}
 \input{testbeam}

\section{Performance in test-beam measurements}
\label{sec:performance_tb}
\input{performance_tb}

\section{Conclusions}
\label{sec:summary}
\input{summary}


\section*{CRediT authorship statement}
\label{sec:credit}
\input{credit_statement}

\section*{Acknowledgments}
\label{sec:acknowledgements}
\input{acknowledgements}

\bibliography{bibliography}

\end{document}

%% file: introduction.tex
The CLIC tracker detector (CLICTD) is a monolithic high-resistivity (HR) CMOS sensor targeting the requirements of the tracking detector for a future Higgs factory such as the Compact Linear Collider (CLIC)~\cite{CLIC_2018_Summary}. 
CLIC is a concept for a linear electron position collider with centre-of-mass energies between 380\,GeV and 3\,TeV. 
The tracking detector of CLIC is subject to stringent requirements~\cite{Dannheim:2673779}. 
A single point resolution of $<\SI{7}{\micro m}$ in one spatial direction needs to be combined with a hit time resolution of $\lesssim 5$\,ns, a hit detection efficiency $>\SI{99.7}{\percent}$ and a maximum material budget of 1--2\SI{}{\percent} $X_0$ per detector layer. 

Monolithic CMOS silicon sensors are attractive candidates for the large-area CLIC tracking detector due to their small material budget and relative ease of large-scale production.
It is advantageous to minimise the input capacitance of these sensors to profit from a low noise level, a low detection threshold, a high signal, and a low power consumption~\cite{Snoeys:2013bka}. 
A small capacitance can be achieved by minimising the size of the collection diode~\cite{Senyukov:2013se}. 
Various design features of the 180\,nm CMOS imaging process have been successfully tested within the framework of the ALPIDE sensor development for the ALICE Inner Tracking System upgrade~\cite{AglieriRinella:2017lym, Dannheim:2019myr}.
However, fast charge collection is hampered by the limited depletion and the low electric field in the small collection electrode design.
To mitigate this, modifications to the sensor design have been introduced in order to achieve full lateral depletion of the epitaxial layer~\cite{SNOEYS201790} and to  enhance the lateral field for additional acceleration of the charge collection~\cite{Munker_2019}.
The optimised sensor designs have been shown to improve radiation hardness in the Mini-MALTA sensor developed in the context of the ATLAS upgrade Phase-II~\cite{Dyndal:2019nxt}. 
The CLICTD sensor is fabricated in two different pixel flavours affecting the charge collection. 
In this document, the measurements in charged particle beams are presented and the performance of the two pixel flavours is compared.

%% file: clictd.tex
The CLICTD sensor features a matrix of $16\times128$ detection channels with a size of \SI{300x30}{\micro m}.
In the \SI{300}{\micro m} column dimension, the channels are segmented into eight sub-pixels, each with its own collection diode and analogue front-end.
This segmentation scheme reduces the digital footprint while maintaining fast charge collection and a small detector capacitance.     
In the following, the sensor and readout design are outlined.
A detailed description is presented elsewhere~\cite{clictd_design_characterization}.

\subsection{Sensor design}

\begin{figure*}
	\centering
	\begin{subfigure}{.5\textwidth}
		\centering
		\includegraphics[width=.99\linewidth]{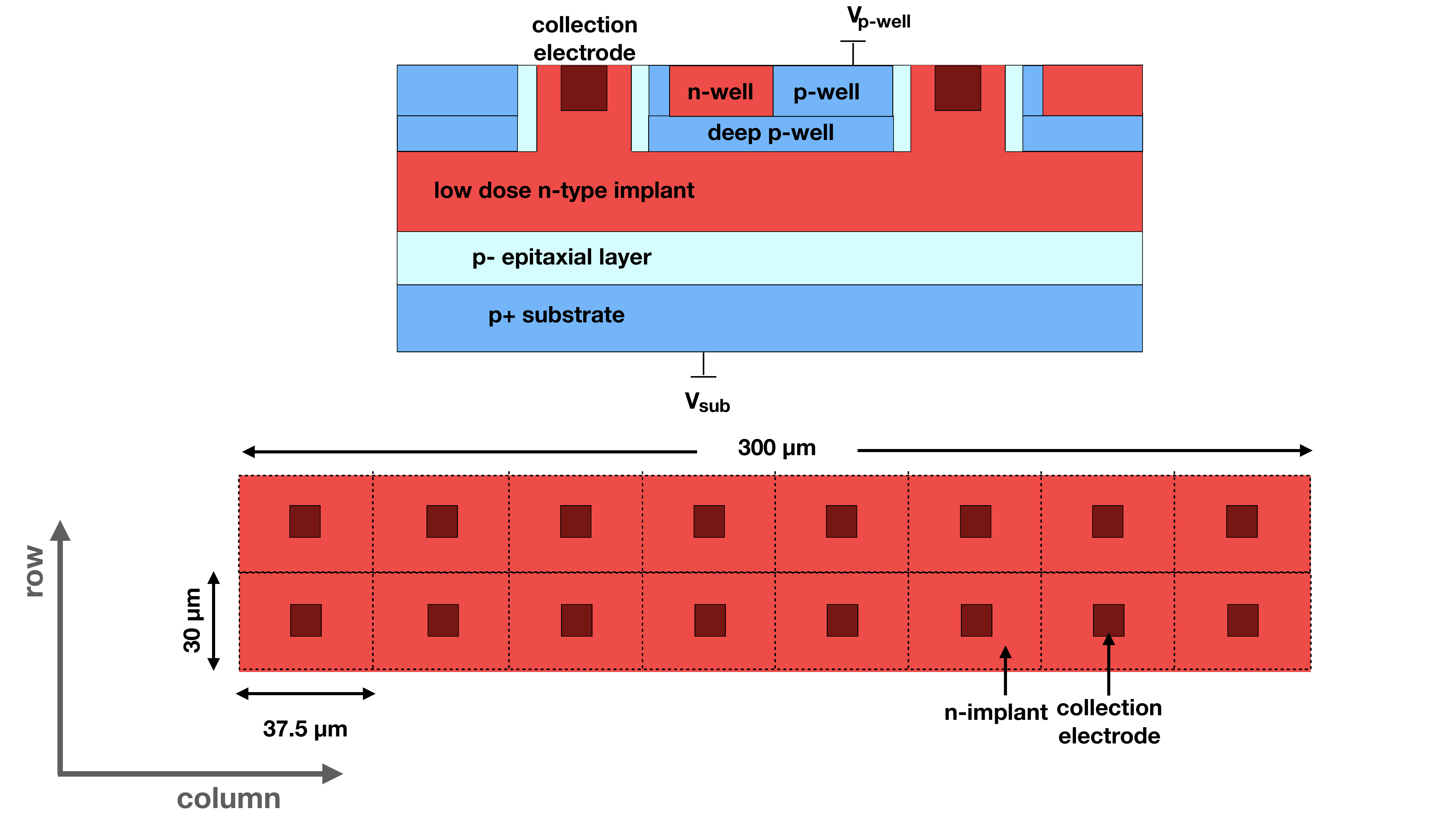}
		\caption{Continuous n-implant}
		\label{fig:sensor_designs_a}
	\end{subfigure}%
	\begin{subfigure}{.5\textwidth}
		\centering
		\includegraphics[width=.99\linewidth]{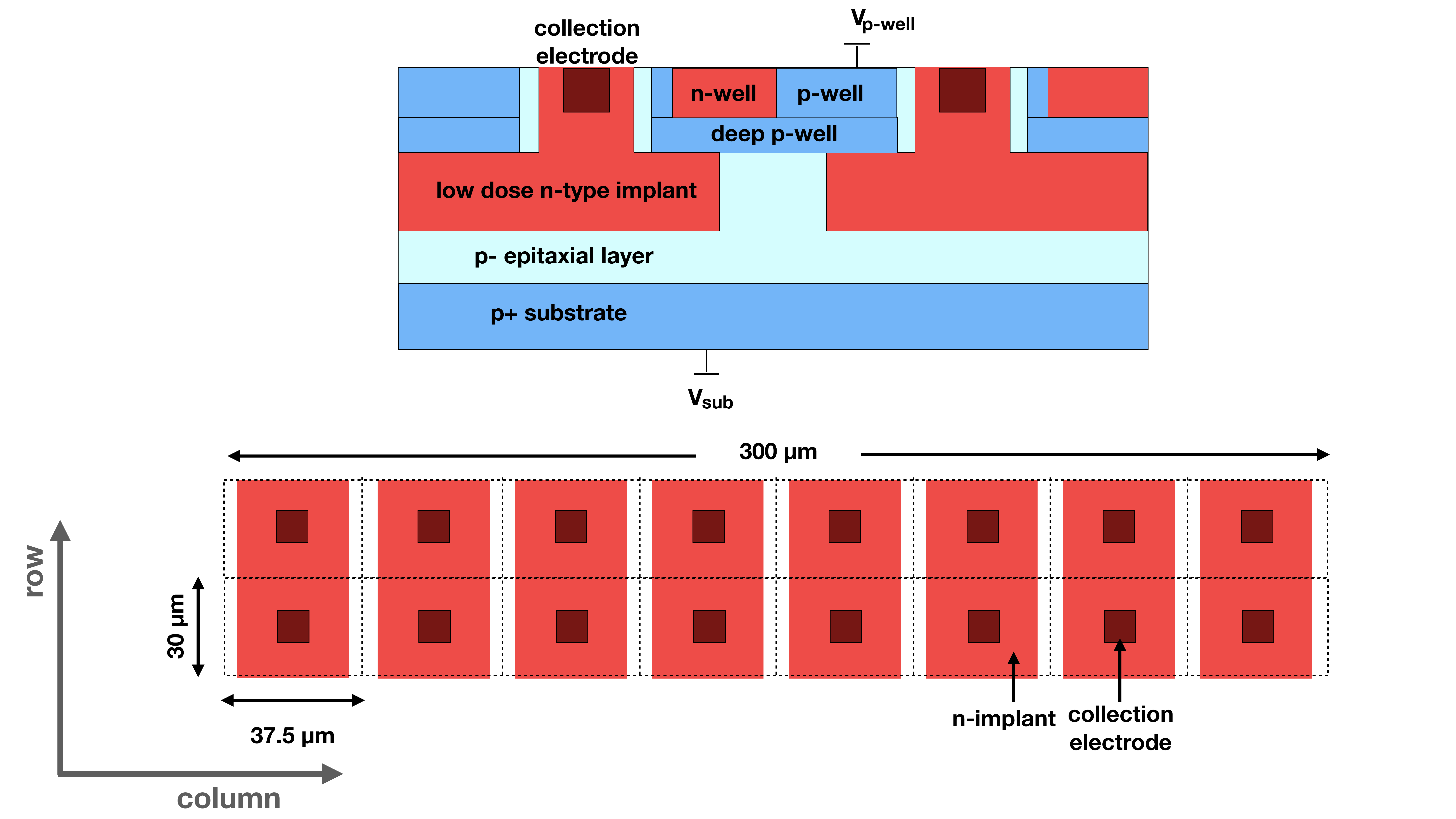}
		\caption{Segmented n-implant}
		\label{fig:sensor_designs_b}
	\end{subfigure}
	\caption{Sensor designs}
	\label{fig:sensor_designs}
\end{figure*}

The CLICTD sensor is fabricated in a 180\,nm CMOS imaging process~\cite{SNOEYS201790}.
The sensor layout is shown schematically in Fig.~\ref{fig:sensor_designs}. 
The sensor features a small n-type collection electrode placed on a high resistivity p-type epitaxial layer with a thickness of $\SI{30}{\micro m}$. 
The epitaxial layer is grown on top of a p-type bulk substrate, resulting in a 
total thickness of $\SI{300}{\micro m}$ for the entire sensor.
Samples with a total thickness of $\SI{100}{\micro m}$ and $\SI{50}{\micro m}$ have been produced by backside grinding. 
The analogue and digital on-channel electronics are located on deep p-wells, which shield the CMOS transistors from the electric field in the sensor. 
Moreover, the sensor is shielded from the circuitry, which could act as a noise source.  The shielding is also necessary to avoid charge collection by electrodes other than the collection electrode. 

From 3D TCAD simulations, it is expected that the depletion zone extends approximately \SI{23}{\micro m} in depth.
A full lateral depletion in the epitaxial layer can be achieved by including a low-dose n-type implant underneath the p-wells. 
In a second pixel flavour, a segmentation of the implant is introduced in order to speed up charge collection~\cite{Munker_2019}, as illustrated in Fig.~\ref{fig:sensor_designs_b}.
The segmented n-type implant generates a lateral electric field, resulting in a faster propagation of charge carriers to the collection diodes. 

The CLICTD sensor is fabricated in both pixel variants: the first flavour with a continuous n-implant and the second flavour with a segmented n-implant.  
For the second flavour, the segmentation is only applied along the column dimension of the matrix, as indicated schematically on the bottom of Fig.~\ref{fig:sensor_designs_b}. 
In the row dimension, which would be perpendicular to the magnetic field in the CLIC detector, the implant is not segmented since charge sharing is desired to improve the spatial resolution. 

Reverse biases are applied to the substrate and to the p-wells. 
The  p-well reverse bias is limited to -6\,V to avoid a breakdown of the on-channel NMOS transistors~\cite{vanHoorn:2119197}.
The difference between substrate and p-well bias is limited as well to avoid punch-through between them~\cite{Munker:2019vdo}.

\subsection{Analogue and digital front-end}

The analogue front-end in each sub-pixel features a voltage amplifier connected to a discriminator, where the voltage pulse is compared to an adjustable detection threshold.
Variations of the effective threshold in each sub-pixel can be corrected with a 
a 3-bit threshold-tuning DAC. 
The discriminator outputs of the eight sub-pixels in each detector channel are combined in the on-channel digital front-end with an \textit{OR} gate.
Each detector channel records the binary hit pattern of its eight sub-pixels.  

For timing measurements, the time-of-arrival (ToA) is recorded with a 100\,MHz clock.
A global shutter signal sets the time reference for the ToA, which is defined as the number of clock cycles from crossing the threshold until the shutter is closed.
The granularity of the ToA sets a lower limit of $\SI{10}{ns}/\sqrt{12} = 2.9$\,ns to the measured time resolution.
Noise and jitter in the amplifier originating e.g. from the input clock and the logic can degrade the front-end time resolution further.
For energy measurements, the time-over-threshold (ToT) is recorded, which is defined as the number of clock cycles between rising above and falling below the detection threshold.

When the sub-pixel discriminator outputs are combined, the ToA is set by the sub-pixel with the earliest timestamp.
The ToT corresponds to the number of clock cycles during which at least one sub-pixel is above threshold.
This readout scheme limits the achievable measurement precision if more than one sub-pixel detects a signal above threshold, due to the ambiguity of the ToA and ToT assignment in this case. 

%% file: assemblies.tex
The results described in the following have been obtained with the Caribou versatile data acquisition system~\cite{Vanat:2020eul}.
All the reported measurements were obtained from a single $\SI{300}{\micro m}$ thick assembly for each pixel flavour, except for those in Section~\ref{sec:performance_efficiency} where samples of each flavour with thicknesses of $\SI{50}{\micro m}$ and $\SI{100}{\micro m}$ were used to compare the efficiency.  
The sensor bias voltage at the p-wells/substrate is fixed to -6\,V/-6\,V.
The effect of a lower absolute bias voltage is presented elsewhere~\cite{dort2020clictd}. 

%% file: configurations.tex
Initial laboratory measurements for the CLICTD sensor can be found elsewhere~\cite{clictd_design_characterization}. 
In this document, measurements crucial for the calibration and reconstruction of the test-beam data are summarised and partially extended.
The results are summarised in Table~\ref{tab:lab_results}.

\begin{table*}[t]
	\centering
	\begin{tabular}{lll}
		\hline
		\toprule
		\textbf{Parameter} & \textbf{Continuous n-implant} & \textbf{Segmented n-implant} \\ 
		\midrule
		Conversion factor [e$^-$/DAC] & $8.99 \pm 0.02 \textrm{ (stat.)} ^{+0.20}_{-0.27} \textrm{ (syst.)}$ & $8.93 \pm 0.02 \textrm{ (stat.)}  ^{+0.20}_{-0.27}  \textrm{ (syst.)}$ \\
		Threshold dispersion [e$^-$]& $24 \pm 3 \textrm{ (syst.)}$  &  $24 \pm 3 \textrm{ (syst.)}$ \\
		Single pixel noise [e$^-$] & $11 \pm 1 \textrm{ (syst.)}$ & $11 \pm 1 \textrm{ (syst.)}$  \\
		Operation threshold (test-beam) [e$^-$] & $170 ^{+ 4}_{- 5} \textrm{ (syst.)}$ & $178 ^{+ 4}_{- 5}  \textrm{ (syst.)}$ \\ 
		Front-end time resolution [ns] & $5.1  \pm 0.1 \textrm{ (stat.)} \pm 0.1 \textrm{ (syst.)}$ & $5.1 \pm 0.1 \textrm{ (stat.)} \pm 0.1 \textrm{ (syst.)}$ \\ 
		\bottomrule
	\end{tabular}
	\caption{Results of the laboratory characterisation for the pixel flavour with continuous n-implant and segmented n-implant.
	The statistical uncertainty is marked by (stat.), the systematic uncertainty by (syst.).}
	\label{tab:lab_results}
\end{table*}

\subsection{Threshold calibration}

The detection threshold is calibrated using the X-ray fluorescence spectra of up to four different materials.
A linear relationship between the threshold DAC values and energy is found.

Two sources of systematic uncertainties are studied: 
First, the analysis is repeated with different numbers of materials and varied fit ranges. 
A maximum deviation of $\pm 0.2$\,e$^-$/DAC is found. 
Second, a maximum charge collection loss of 30\,e$^-$ due to sub-threshold effects and charge carrier recombination is assumed, which yields a one-sided uncertainty of $- 0.07$\,e$^-$/DAC.
The statistical uncertainty amounts to $\pm 0.02$\,e$^-$/DAC.

The resulting conversion factors are 
$$
8.99 \pm 0.02 \textrm{ (stat.)} ^{+0.20}_{-0.27} \textrm{ (syst.)} \textrm{ e}^-\textrm{/DAC},
$$
and 
$$
8.93 \pm 0.02 \textrm{ (stat.)} ^{+0.20}_{-0.27} \textrm{ (syst.)} \textrm{ e}^-\textrm{/DAC}
$$
for the pixel flavour with continuous and segmented n-implant, respectively.
The sensor capacitance is expected to be unaffected by the pixel flavours.
Therefore, the conversion factors are the same within the uncertainties. 

\subsection{Threshold equalisation }

The sub-pixel threshold variation is reduced using the 3-bit threshold-tuning DAC.
The systematic uncertainty is estimated by propagating the uncertainty on the threshold conversion factor to the threshold dispersion, which yields $^{+ 0.5}_{- 0.8}$\,e$^-$ for both pixel flavours. 
Additional sources of systematic uncertainties are studied by repeating the equalisation with different environmental conditions and readout schemes. 
In total, a maximum deviation of $\pm 3$\,e$^-$ is found.
The statistical uncertainty is in the sub-electron range and therefore negligible.

After equalisation, the RMS of the threshold dispersion amounts to ($24 \pm 3$)\,e$^-$ for both pixel flavours.
The threshold equalisation is repeated with additional CLICTD samples and the results are found to be within the stated uncertainties.

\subsection{Single sub-pixel noise}

The single sub-pixel noise is estimated by varying the detection threshold around the baseline and registering the noise hits for every sub-pixel and every threshold value.
The RMS of the noise hit distribution as a function of the threshold is extracted on sub-pixel level. 

The systematic uncertainty on the average sub-pixel RMS is estimated with the same techniques used for the calculation of the threshold dispersion uncertainty and a value of $\pm 1$\,e$^-$ is found. 
The statistical uncertainty is below one electron.

The mean of the single sub-pixel noise RMS is ($11 \pm 1$)\,e$^-$ for both pixel flavours.

\subsection{Operation  threshold}

The operation threshold is defined as the lowest mean threshold at which a noise-free operation ($< 1 \times 10^{-3}$\,hits/sec) of the sensor is possible.
Sub-pixels exhibiting non-Gaussian noise are excluded from this definition by masking them online. 
The number of masked pixels is less than one per mille of the entire matrix. 

The systematic uncertainty is calculated by propagating the uncertainty in the threshold conversion factor to the operation threshold.
The statistical uncertainty with a value below one electron is negligible.

With this definition, the operation threshold in the laboratory is found to be $135^{+ 4}_{- 5}$\,e$^-$.
In the test-beam, the operation threshold has been increased to  $170^{+ 4}_{- 5}$\,e$^-$ for the pixel flavour with continuous n-implant and $178^{+ 4}_{- 5} $\,e$^-$ for the flavour with segmented n-implant in order to gain margin for stable operation. 

\subsection{ToT calibration}

The ToT measurement is calibrated with test-pulses injected into the analogue front-end amplifier of individual sub-pixels.
The relationship between the known amplitude of the test-pulses and the measured ToT is parametrised with a non-linear function depending on four parameters~\cite{Jakubek:2011dsm}.

The ToT calibration has been shown to have limited precision for the following reasons:
\begin{itemize}
	
	\item The capacitance of the test-pulse injector is insufficient to trigger the highest possible ToT in all sub-pixels. 
	Some of the sub-pixels are therefore not calibrated correctly in the high-ToT range.
	This constraint affects single pixel signals with $\gtrsim 2.1$\,ke$^-$.
	\item The on-channel NMOS transistors are affected by the negative bias voltage applied to the p-wells. 
	As a consequence, the operation margin of the circuits is reduced leading to a strong non-uniformity and non-linearity in the ToT response~\cite{clictd_design_characterization}. 
	\item In cases where several sub-pixels are hit in the same readout frame, the combined ToT response is assigned to all hit sub-pixels. 
	This can lead to a wrong conversion depending on the observed hit multiplicity. 

\end{itemize}
 
The ToT calibration is applied to evaluate the signal in Section~\ref{sec:performance_tb}.
For all other analyses, the ToT values are not converted to electrons to prevent the introduction of systematic errors due to the above mentioned limitations. 

The insight acquired from the characterisation of the CLICTD front-end will have important implications for the front-end development of the next generation of sensors. 

\subsection{Time resolution of front-end circuits}

The time resolution of the front-end is estimated with test-pulses injected in the analogue and digital front-end of each sub-pixel.
The test-pulse injection is triggered approximately \SI{1}{\micro s} before the end of a CLICTD frame and occurs asynchronously to the ToA clock in order to ensure a random phase between the clock and the injection time.  
For each sub-pixel, the time residuals between the ToA values recorded for the analogue and digital test-pulse are evaluated.
To estimate a lower limit on the front-end time resolution, only pulse heights that induce a ToT of 11 are considered, which is equivalent to the most probable energy deposition of a minimum ionising particle.

The systematic uncertainty introduced by the uncertainty on the threshold conversion factor is negligible. 
The analysis is repeated by lowering and raising the considered signal range by one ToT bin and a deviation of $\pm 0.1$\,ns is found. 
The statistical uncertainty is $\pm 0.1$\,ns. 

For both pixel flavours, the measurement yields an RMS of the ToA residual distribution of:
$$
5.1 \pm 0.1 \textrm{ (stat.)} \pm 0.1 \textrm{ (syst.)}\textrm{ ns}.
$$
The front-end time resolution is therefore larger than estimated from the 10\,ns ToA binning.

%% file: testbeam.tex
In the following, the test-beam setup and the offline reconstruction are outlined. 

\subsection{Beam telescope}

Data was recorded at the DESY II Test Beam Facility~\cite{Diener:2018qap} using a 5.4\,GeV electron beam.

\begin{figure}[tbp]
	\centering
	\includegraphics[width=\columnwidth]{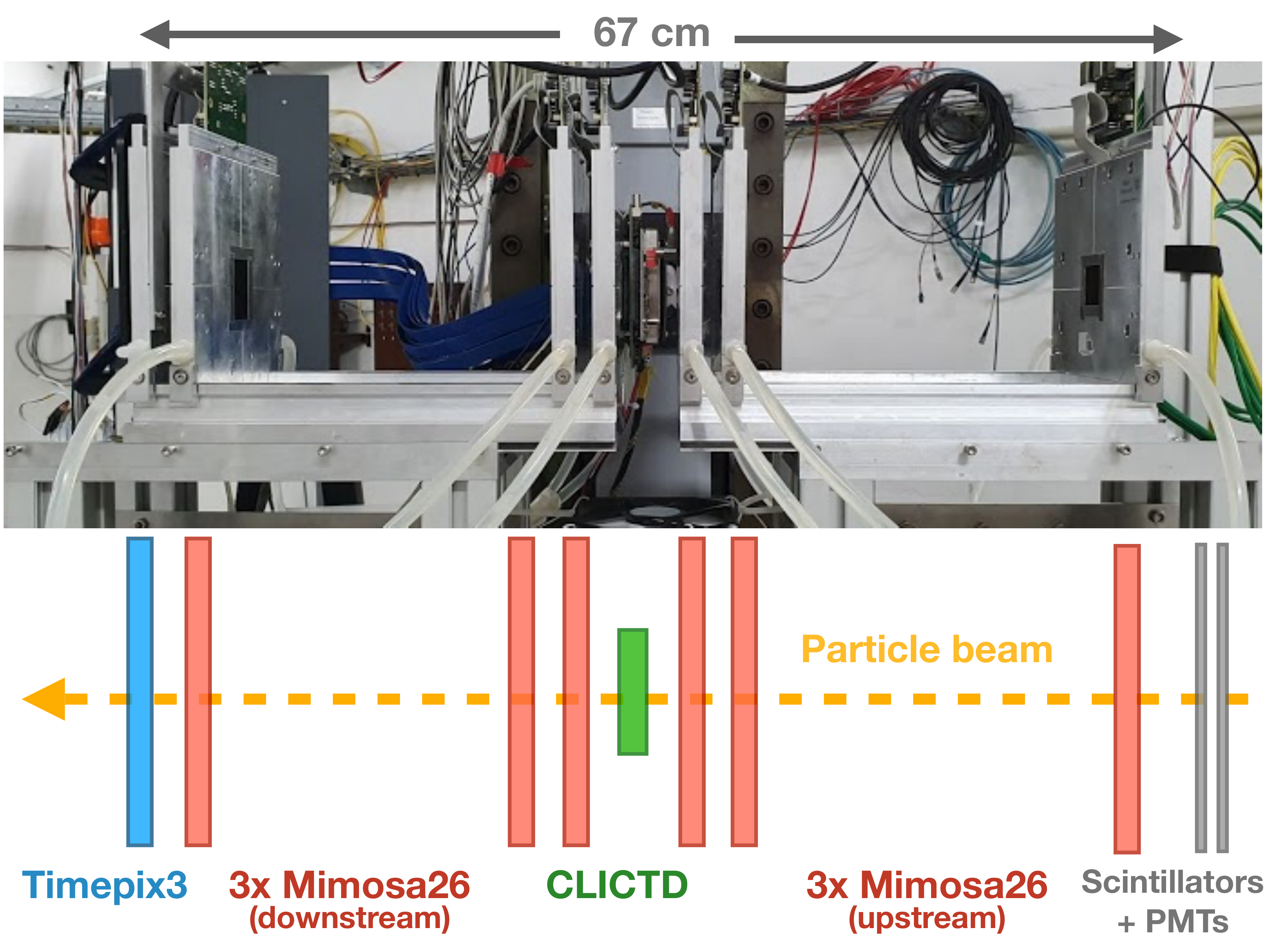}%
	\caption{Telescope setup at the DESY II test-beam facility. The DUT is placed between three MIMOSA-26 planes in the upstream and downstream arm, respectively. An additional Timepix3 plane is placed behind the last MIMOSA-26 plane in the downstream arm.}
	\label{fig:mimosa_telescope}
\end{figure}

For the measurements presented in this document, the device under test (DUT) is placed in a EUDET telescope with six MIMOSA-26 monolithic active pixel sensors~\cite{Jansen:2016bkd} and a Timepix3 time-reference plane~\cite{Poikela_2014}, as illustrated in Fig.~\ref{fig:mimosa_telescope}.

The AIDA Trigger Logic Unit (TLU)~\cite{Baesso:2019smg} provides a trigger signal when recording a coincidence between two scintillators placed in front of the first telescope plane.
The telescope and DUT are controlled and read out using the  EUDAQ2 data acquisition framework~\cite{Liu:2019wim}.

The CLICTD sensor is operated with shutters closing 200\,ns after a trigger signal is received from the TLU.
The shutter is opened again after the readout is completed. 
A compression algorithms ensures that only data coming from those detector channels that registered a hit are shifted out.
The resulting readout frequency is approximately 500\,Hz.

\subsection{Reconstruction and analysis of test-beam data}

The offline analysis of the test-beam data is performed with the test-beam reconstruction framework Corryvreckan~\cite{corry_paper, corry_manual}. 

The offline event building is based on the readout frames provided by the CLICTD sensor. 
For the MIMOSA planes, only pixel hits that are associated to a TLU trigger signal with a timestamp within the CLICTD frame are considered.
Likewise, Timepix3 pixel hits are required to have a timestamp that lies within a CLICTD frame.
This event building scheme ensures that the telescope data was recorded when the DUT was active. 

For each telescope plane, adjacent pixels hits are combined into \textit{clusters}.
The cluster position is calculated by a centre-of-gravity algorithm.

Track candidates are selected by requiring a cluster on each of the seven telescope planes. 
The tracks are fitted with the General Broken Lines (GBL) formalism~\cite{Blobel:2006yi}, which takes into account multiple scattering in the material traversed by the beam particles. 
In the alignment procedure of the telescope planes, the track $\chi^2$ is minimised.
For the subsequent analysis, only tracks with a $\chi^2$ per degree of freedom less than or equal to five are considered. 

For the measurements in this document, the resolution of the track impact position on the DUT is determined to be between $\SI{2.4}{\micro m}$ and $\SI{2.8}{\micro m}$, depending on the telescope plane spacing~\cite{Jansen:2016bkd, resolution_simulator}. 
The timestamp of each track is given by the ToA measurement in the Timepix3 plane. 
The track time resolution is determined with the methods described in~\cite{Pitters:2019yzg} and a value of 1.1\,ns is found.

A reconstructed track is associated to a CLICTD cluster if the spatial distance between the track intercept position on the DUT and the nearest pixel in a cluster is less than 1.5 pixel pitches in both directions and the track time is within the same CLICTD frame as the cluster.  

The hit detection efficiency of the DUT is defined as the number of tracks associated to a CLICTD cluster over the total number of tracks.
The tracks used for the hit detection efficiency calculation are required to pass through the acceptance region of the DUT.
The acceptance region comprises the physical pixel matrix of the sensor excluding one column/row at the matrix edge.
It has been verified that the exclusion of additional columns/rows close to the matrix edge does not alter the results.
If the track intercept position on the DUT lies within a masked pixel or its direct neighbours, the track is rejected. 
The efficiency uncertainty is calculated using a Clopper-Pearson confidence interval of one sigma~\cite{clopper_pearson}. 

\begin{figure}[tbp]
	\centering
	\includegraphics[width=\columnwidth]{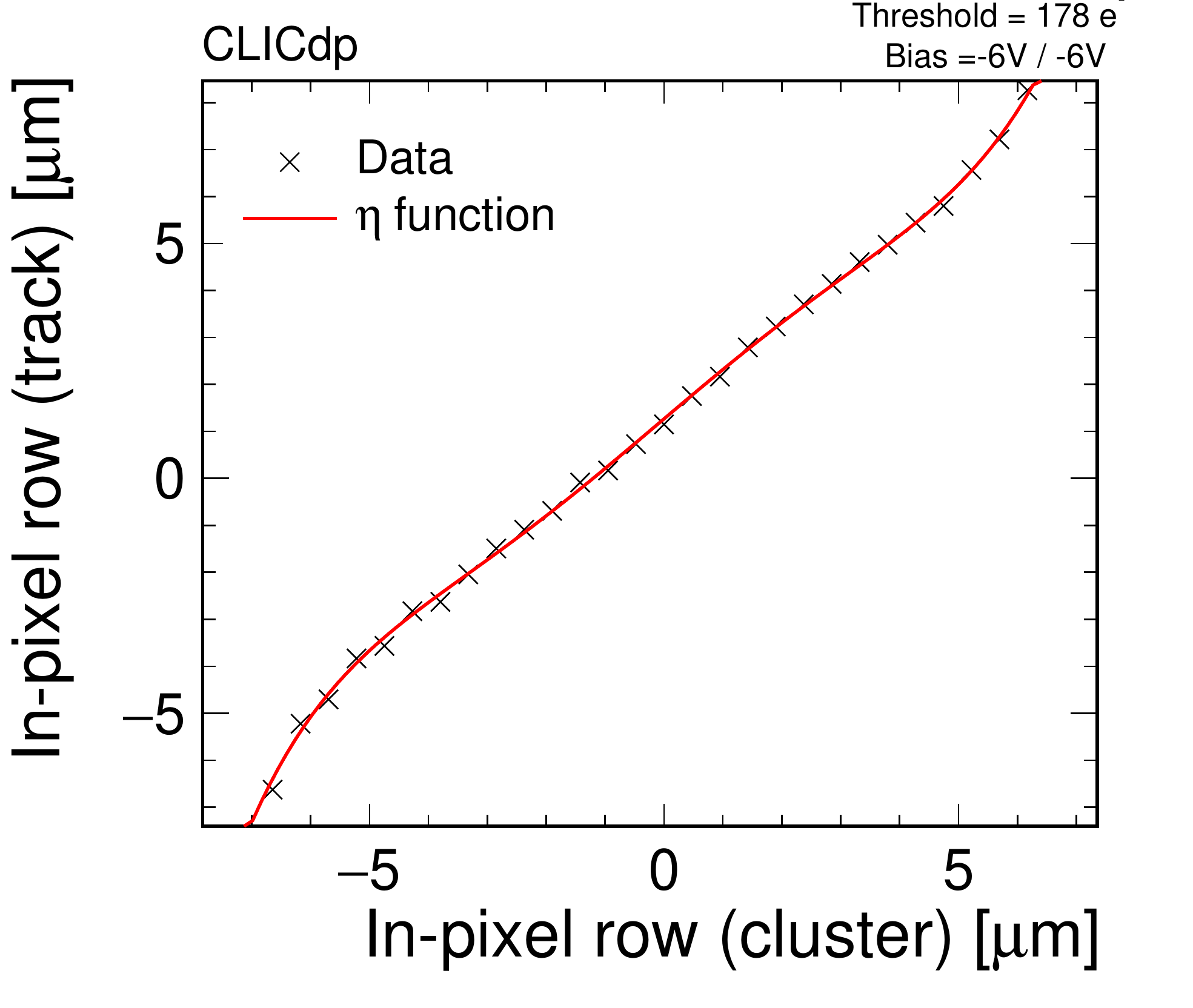}%
	\caption{In-pixel telescope track position as a function of reconstructed in-pixel cluster position for the pixel flavour with segmented n-implant at a threshold of 178\,e. 
	The 5th order polynomial fitted to the distribution (shown as red line) is defined as $\eta$-function.}
	\label{fig:eta_function}
\end{figure}

For the CLICTD sensor, the cluster position is corrected with the $\eta$-formalism in order to account for non-linear charge sharing~\cite{Akiba:2011vn}.
The $\eta$-function is constructed by plotting the in-pixel position of the telescope tracks as a function of the reconstructed in-pixel cluster position, as shown in Fig.~\ref{fig:eta_function} for a sample with segmented n-implant at a threshold of ~180\,e. 
A 5th order polynomial fitted to the distribution is used to correct the reconstructed cluster position.
The $\eta$-correction is only applied to clusters that have an extent of two pixels in row direction. 

%% file: performance_tb.tex
\begin{figure}[tbp]
	\includegraphics[width=\columnwidth]{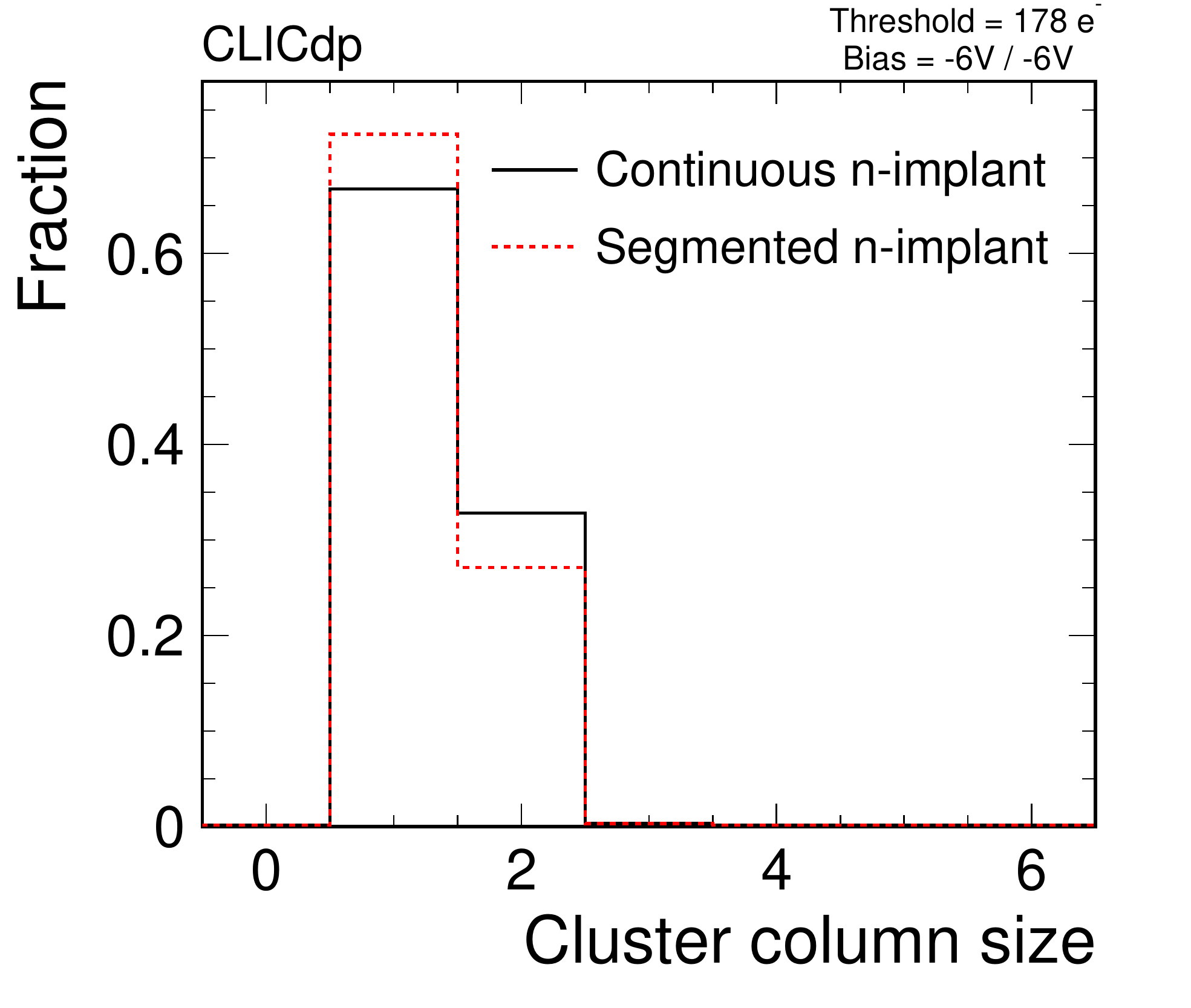}%
	\caption{Cluster size in column direction for the pixel flavour with and without segmentation of the n-implant at nominal conditions.
	The error bars reflecting the statistical uncertainty are not visible.}
	\label{fig:nominal_cluster_size_x}
\end{figure}

\begin{figure}[tbp]
	\centering
	\includegraphics[width=\columnwidth]{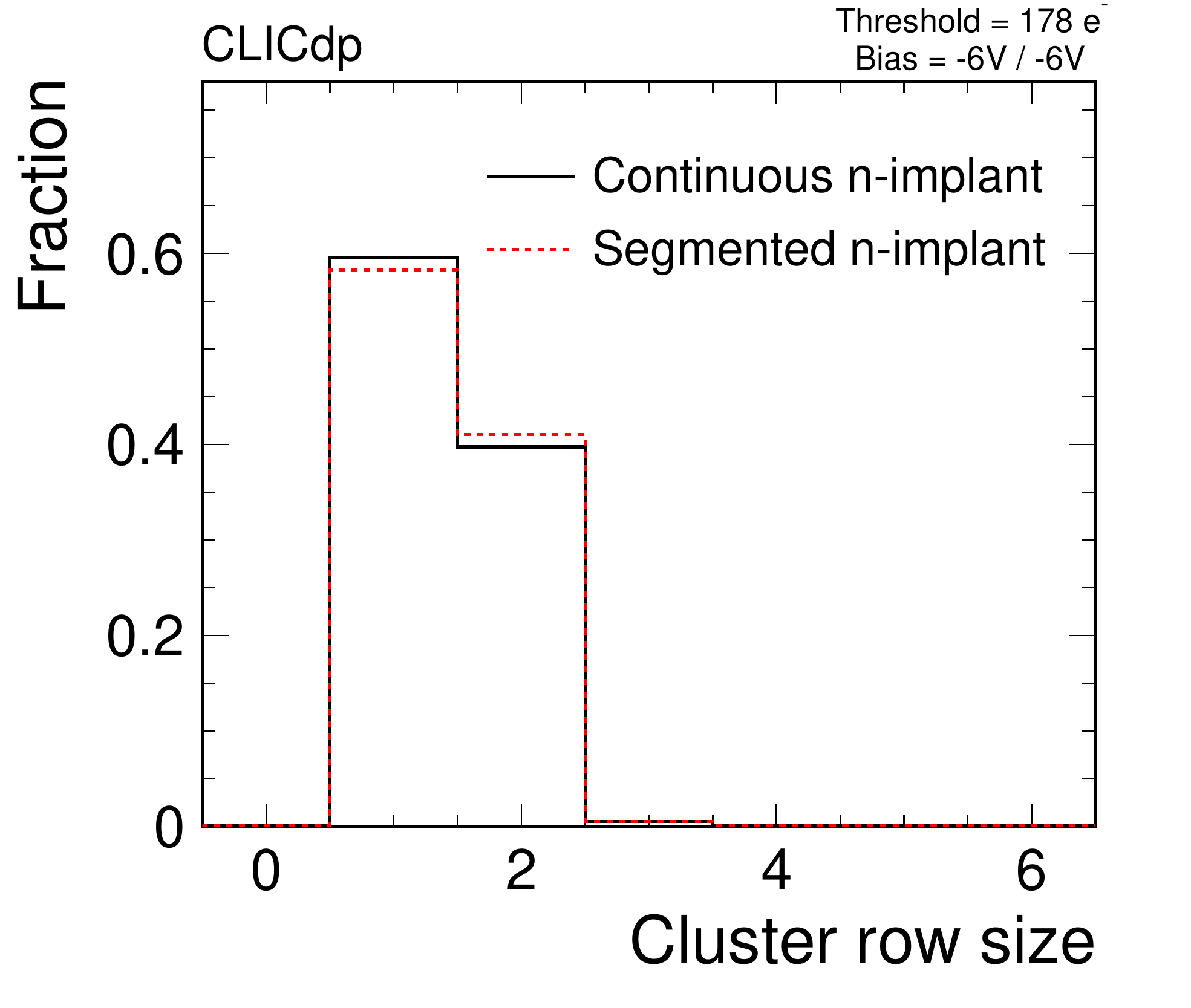}%
	\caption{Cluster size in row direction for the pixel flavour with and without segmentation of the n-implant at nominal conditions.
	The error bars reflecting the statistical uncertainty are not visible.}
	\label{fig:nominal_cluster_size_y}
\end{figure}

In this section, the characterisation of the CLICTD sensor in test-beam measurements is presented. 
First, the performance is evaluated at fixed operating conditions, also referred to as \textit{nominal conditions}.
For the comparison of the two pixel flavours, the detection threshold is fixed to $178$\,e$^-$ and the bias voltage to -6\,V at the p-wells and the substrate. 
In a second step, the applied detection threshold is varied for both flavours in order to study the impact on the performance parameters.
The results are summarised in Table~\ref{tab:tb_results}.


\begin{table*}[tbp]
	\centering
	\begin{tabular}{lll}
		\toprule
		\textbf{Parameter} & \textbf{Continuous n-implant} & \textbf{Segmented n-implant} \\ 
		\midrule
		Cluster size [pixels] & $1.93 \pm 0.01$ (syst.) &  $1.76 \pm 0.01$ (syst.)  \\
		Cluster column size [pixels] & $1.34 \pm 0.01$ (syst.) &  $1.28 \pm 0.01$ (syst.)  \\
		Cluster row size [pixels] & $1.43 \pm 0.01$ (syst.)  & $1.44 \pm 0.01$ (syst.)  \\
		Max. threshold with efficiency $> 99.7\%$ [e$^-$] & $387 \pm 12 \textrm{ (stat.)} ^{+ 9}_{- 12}  \textrm{ (syst.)}$ &  $537 \pm 20 \textrm{ (stat.)} ^{+ 12}_{- 16}  \textrm{ (syst.)}$ \\
		Efficient operation window [e$^-$] &  $207 \pm 12 \textrm{ (stat.)} ^{+ 5}_{- 7}  \textrm{ (syst.)}$ &  $357 \pm 20 \textrm{ (stat.)} ^{+ 8}_{- 11}  \textrm{ (syst.)}$ \\
		Spatial resolution (column) $\sigma^{(s)}_\textrm{col}$ [\SI{}{\micro m}]& 6.7 $\pm$  0.2 (syst.) & 7.6 $\pm$ 0.2 (syst.) \\
		Spatial resolution (row) $\sigma^{(s)}_\textrm{row}$ [\SI{}{\micro m}]& 4.6  $\pm$ 0.2 (syst.) & 4.6  $\pm$ 0.2 (syst.) \\		
		time resolution $\sigma^{(t)}$ [\SI{}{ns}]& 6.5  $\pm$ 0.1 (syst.) & 5.8  $\pm$ 0.1 (syst.) \\
		 \bottomrule
	\end{tabular}
	\caption{Results of test-beam characterisation for the two pixel flavours.
	The results are obtained at the operation threshold of 178\,e$^-$ except for the efficiency measurements, where a threshold window is listed.}
	\label{tab:tb_results}
\end{table*}

\subsection{Charge sharing}
\label{sec:charge_sharing}

\paragraph{Nominal conditions}

The pixel flavour with the segmented n-implant was designed to speed up charge collection. As a consequence, charge sharing between sub-pixels in a single channel is reduced. 
This is reflected in the cluster size in the column and row direction in Fig.~\ref{fig:nominal_cluster_size_x} and Fig.~\ref{fig:nominal_cluster_size_y}, respectively. 

\begin{figure*}[tbp]
	\centering
	\begin{minipage}[b]{0.49\textwidth}
		\includegraphics[width=\linewidth]{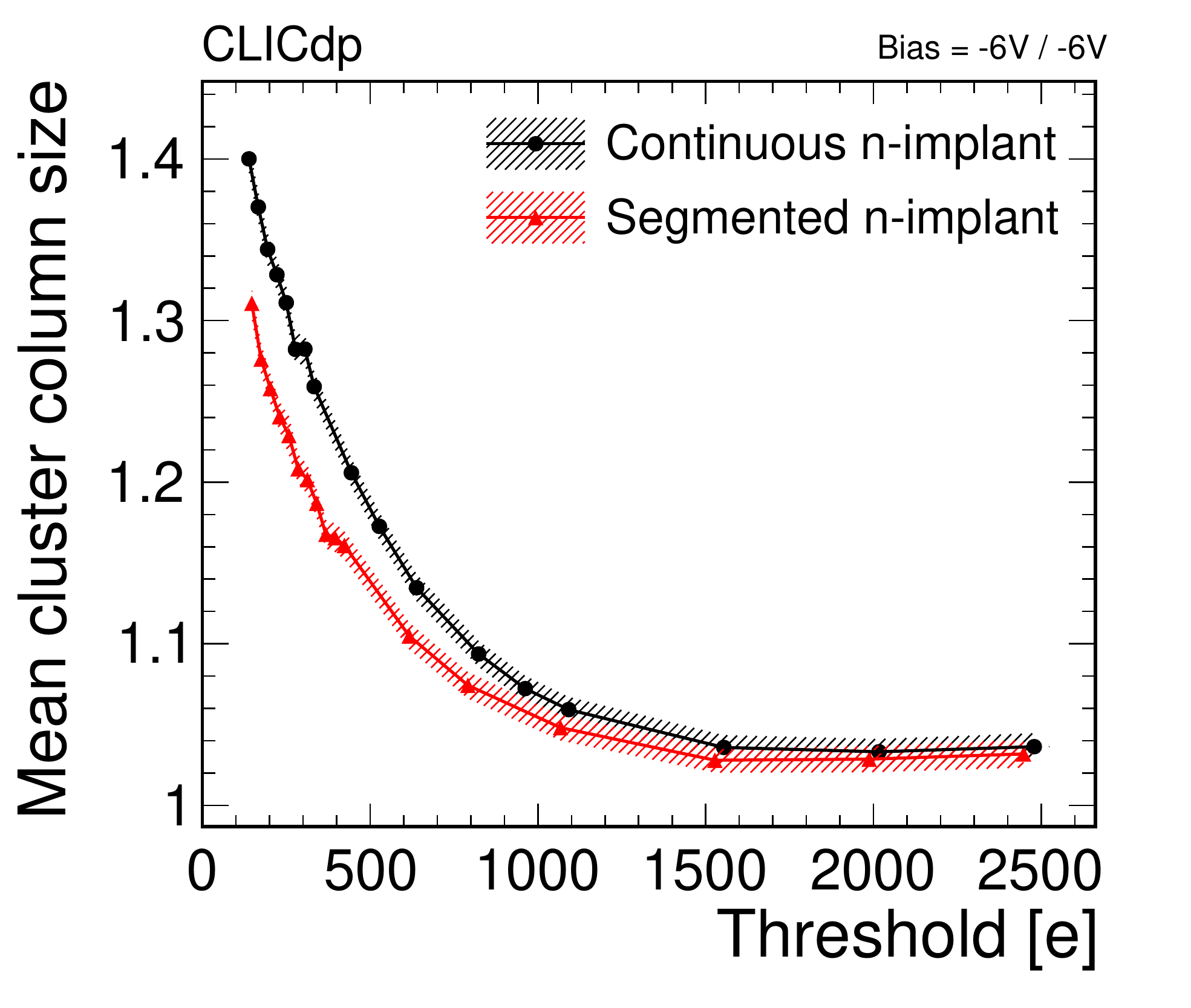}
		\caption{Mean cluster size in column direction as a function of detection threshold.
		The hatched band represents the statistical and systematic uncertainties.}
		\label{fig:cluster_size_scan_x}
	\end{minipage}
	\hfill
	\begin{minipage}[b]{0.49\textwidth}
		\includegraphics[width=\linewidth]{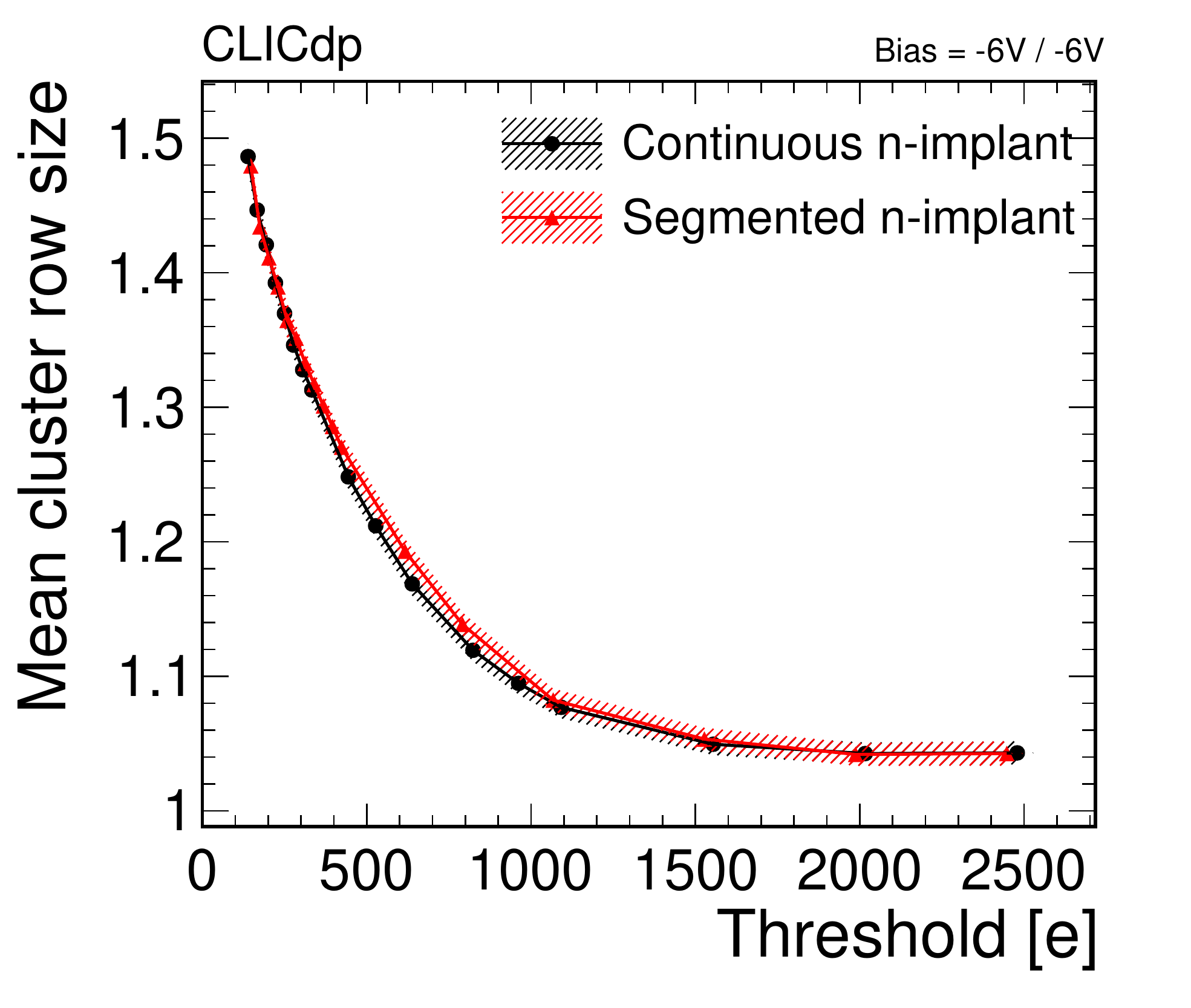}
		\caption{Mean cluster size in row direction as a function of detection threshold.
		The hatched band represents the statistical and systematic uncertainties.}
		\label{fig:cluster_size_scan_y}
	\end{minipage}
\end{figure*}

The statistical uncertainty on the cluster size is of the order of $1 \times 10^{-4}$. 
For the systematic uncertainty, the uncertainty in the threshold conversion factor is propagated to the cluster size using the threshold scans in Fig.~\ref{fig:cluster_size_scan_x} and Fig.~\ref{fig:cluster_size_scan_y}.
A systematic uncertainty of $\pm 0.01$ is found.

The cluster column size is $1.34 \pm 0.01$ and  $1.28 \pm 0.01$ for the pixel flavour with continuous n-implant and segmented n-implant, respectively. 
The reduction by approximately \SI{5}{\percent} for the flavour with the segmented n-implant confirms that charge sharing is suppressed due to the electric field distribution that enforces fast charge collection.
In the row direction, the size is $1.43 \pm 0.01$ and $1.44 \pm 0.01$ for continuous and segmented n-implant, respectively. 
Within the uncertainties, the cluster row size is unaffected by the pixel flavour, confirming that the n-implant segmentation only affects the column direction.  
The results have been cross-checked with several CLICTD samples in order to exclude systematic uncertainties such as a rotational misalignment of the DUT in the test-beam setup. 
The cluster size values agree within the measurement uncertainties.

\begin{figure*}[btp]
	\centering
	\begin{minipage}[b]{0.49\textwidth}
		\includegraphics[width=\linewidth]{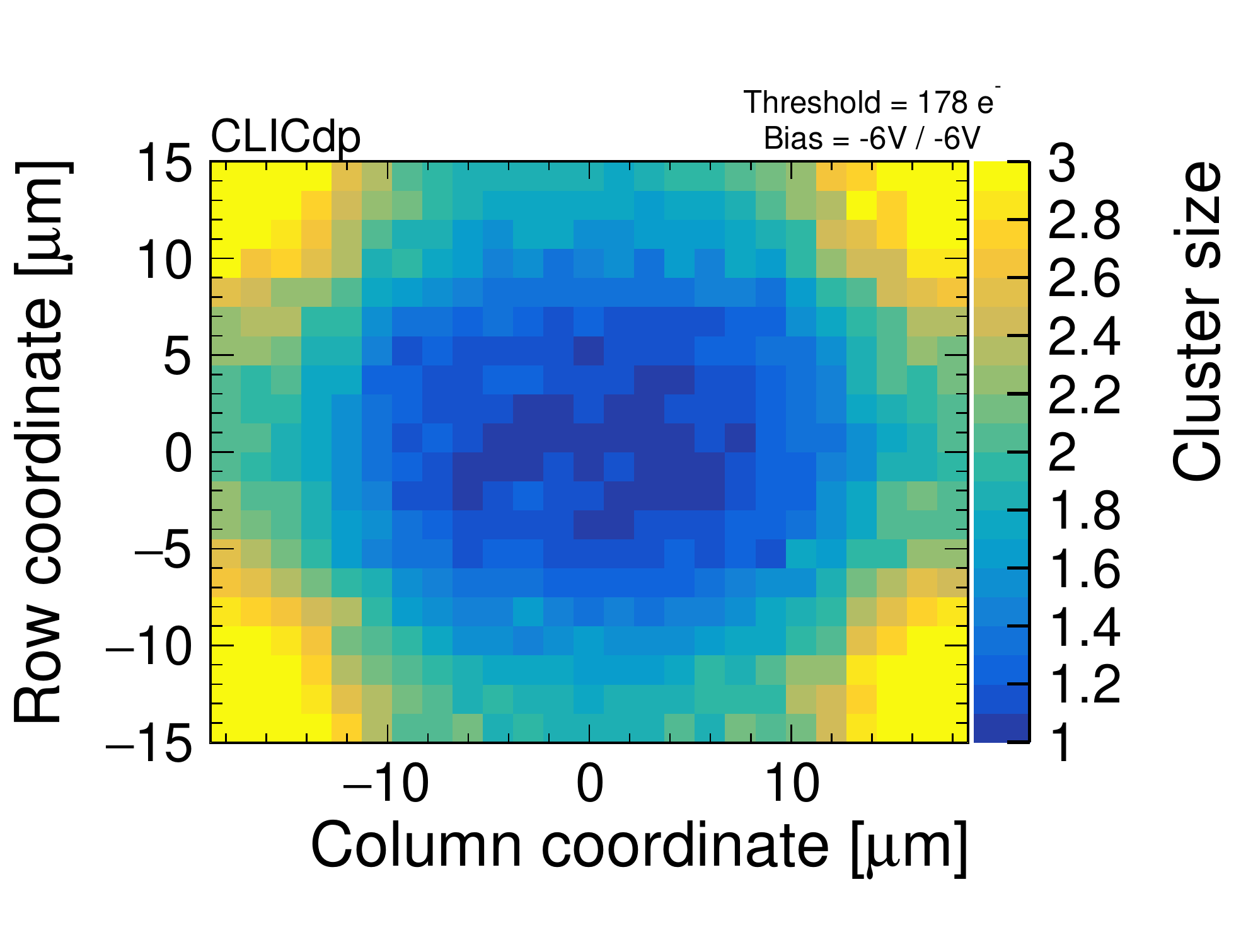}
		\caption{In-pixel cluster size for the flavour with continuous n-implant at nominal conditions.}
		\label{fig:nominal_noGap_cluster_size_inPixel}
	\end{minipage}
	\hfill
	\begin{minipage}[b]{0.49\textwidth}
		\includegraphics[width=\linewidth]{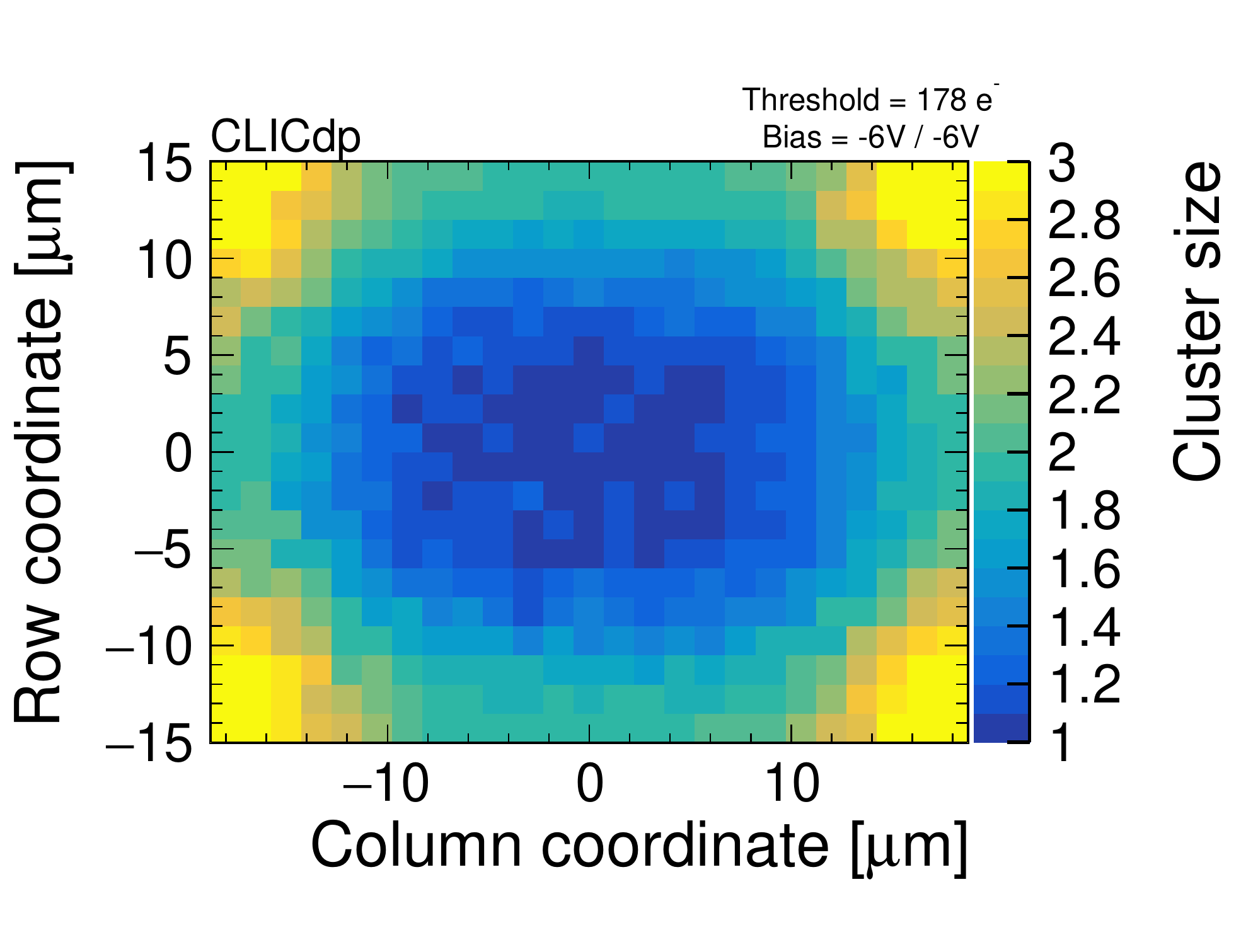}
		\caption{In-pixel cluster size for the flavour with segmented n-implant at nominal conditions.}
		\label{fig:nominal_gap_cluster_size_inPixel}
	\end{minipage}
\end{figure*}

\begin{figure*}[btp]
	\centering
	\begin{minipage}[b]{0.49\textwidth}
		\includegraphics[width=\linewidth]{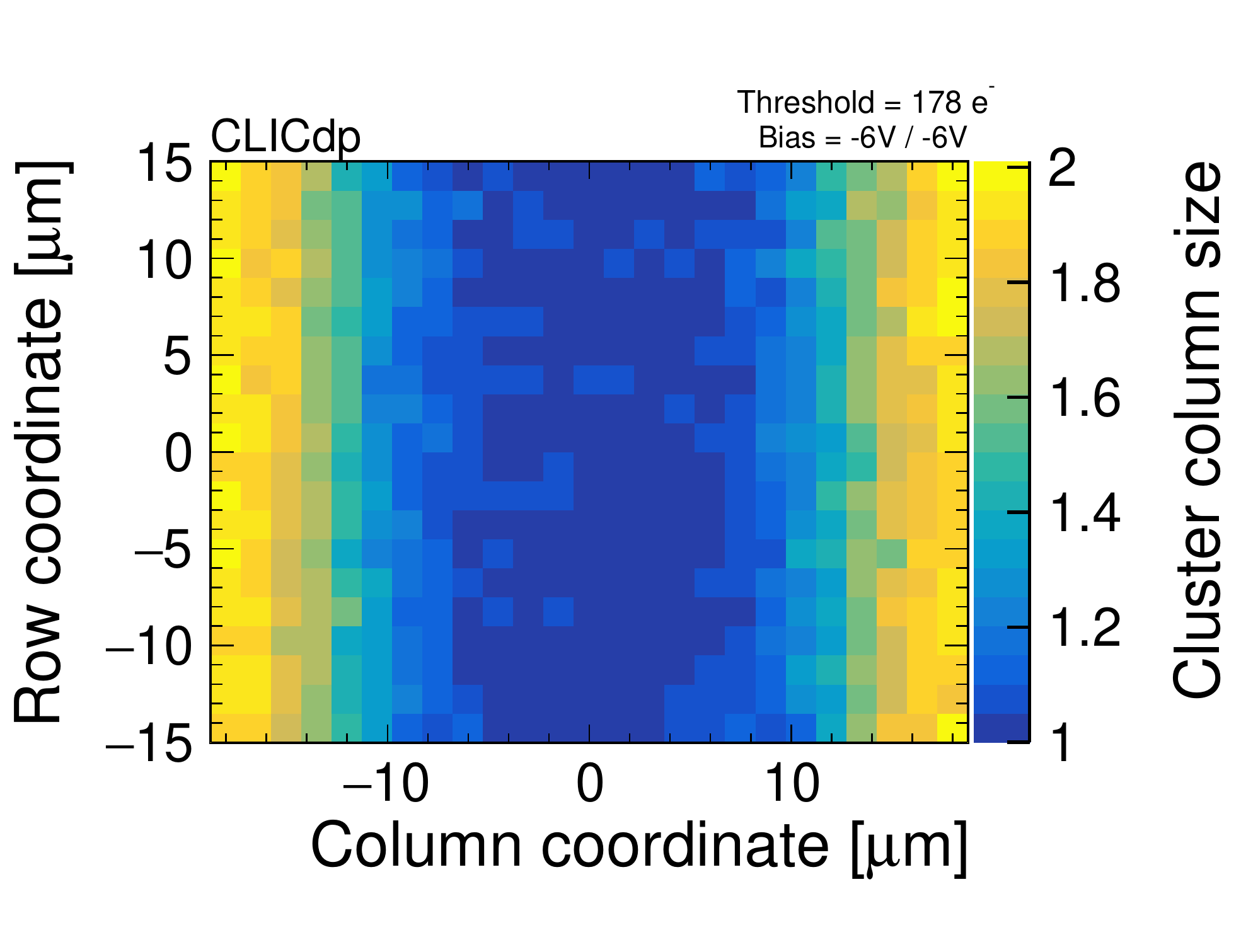}
		\caption{In-pixel cluster column size for the flavour with continuous n-implant at nominal conditions.}
		\label{fig:nominal_noGap_cluster_size_inPixelX}
	\end{minipage}
	\hfill
	\begin{minipage}[b]{0.49\textwidth}
		\includegraphics[width=\linewidth]{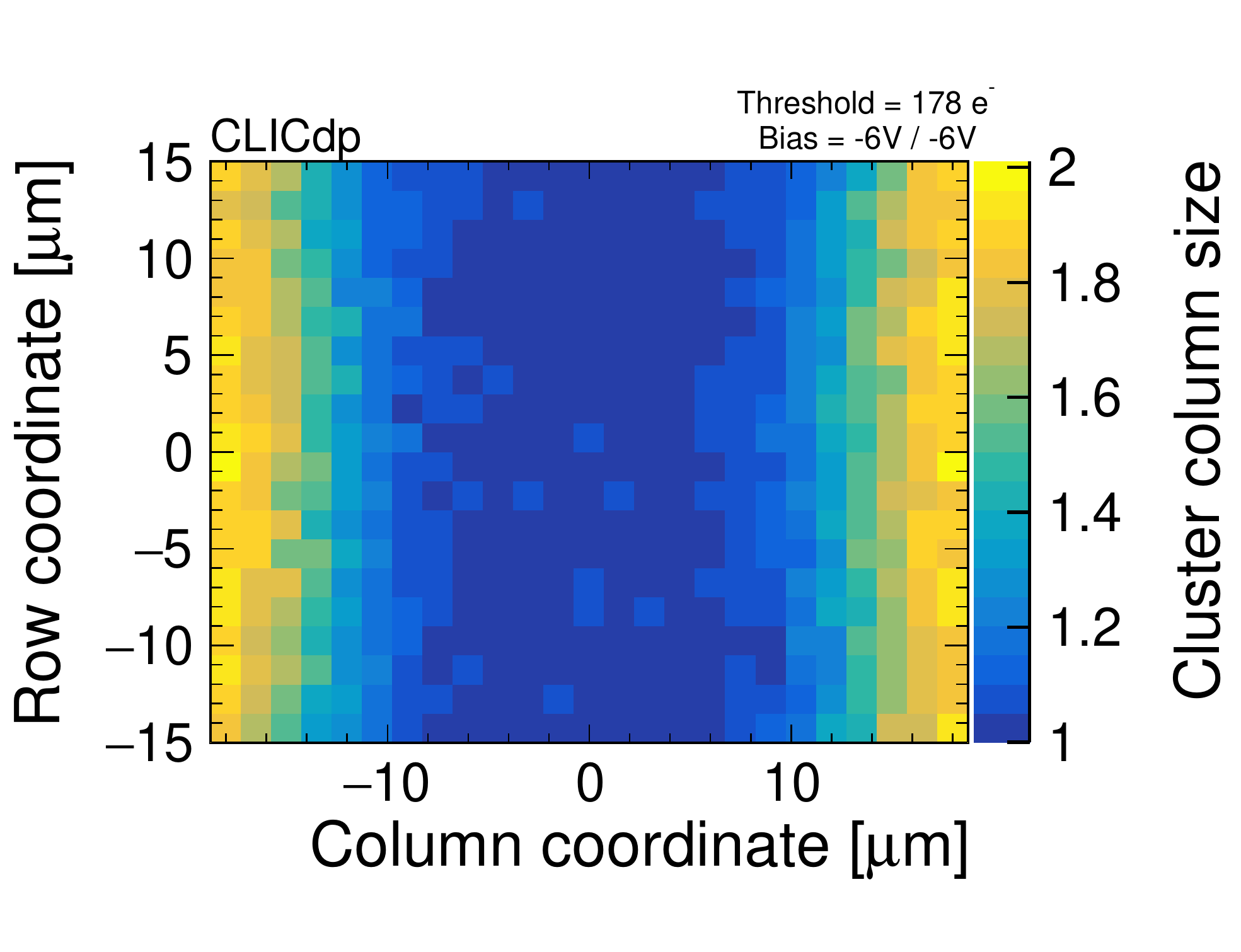}
		\caption{In-pixel cluster column size for the flavour with segmented n-implant at nominal conditions.}
		\label{fig:nominal_gap_cluster_size_inPixelX}
	\end{minipage}
\end{figure*}

Figures~\ref{fig:nominal_noGap_cluster_size_inPixel}--~\ref{fig:nominal_gap_cluster_size_inPixelX} show the cluster size and cluster column size as a function of the in-pixel track intercept position.
The reduction of the cluster size for the pixel flavour with segmented n-implant is indeed only visible in the in-pixel column dimension, where the segmentation is introduced. 

\paragraph{Threshold scan}
With increasing detection threshold, the cluster size decreases, as illustrated in Fig.~\ref{fig:cluster_size_scan_x} and Fig.~\ref{fig:cluster_size_scan_y} for the cluster size in column and row direction, respectively.
In column direction, the impact of reduced charge sharing for the flavour with the segmented n-implant is particularly pronounced for low detection thresholds.
It decreases for high thresholds due to inefficiencies forming at the pixel edges, which are especially sensitive to the different deep n-implant structures.
In the row direction, the size for both pixel flavours is identical over the scanned threshold range.

\subsection{Signal}

\begin{figure}[tbp]
	\centering
	\includegraphics[width=\columnwidth]{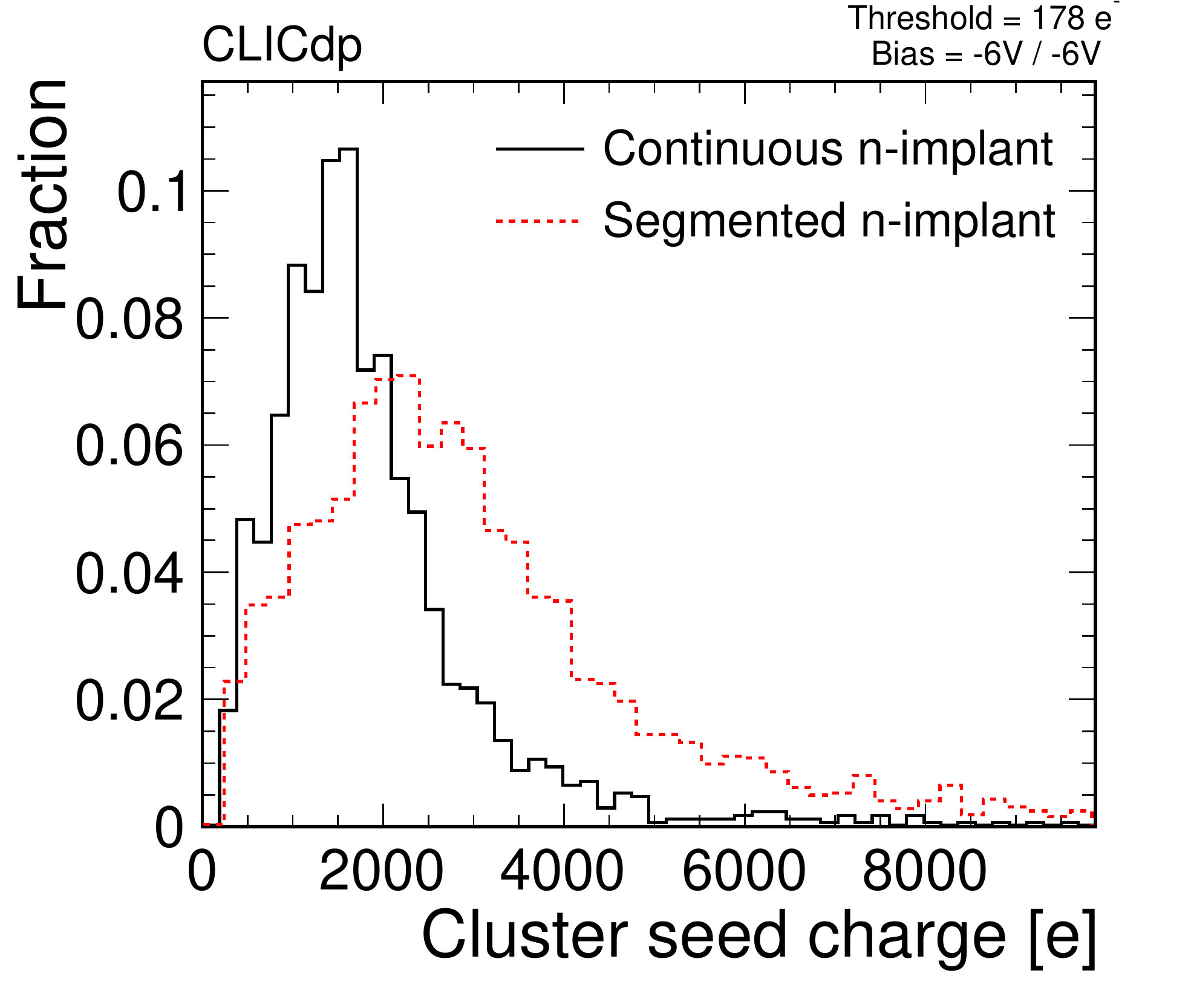}%
	\caption{Cluster seed signal distribution for both pixel flavours at nominal conditions.
	The error bars reflecting the statistical uncertainty are not visible.}
	\label{fig:nominal_cluster_seed_charge}
\end{figure}

The seed signal is defined as the highest single pixel signal in a cluster. 
In Fig.~\ref{fig:nominal_cluster_seed_charge}, the seed signal distribution for both pixel flavours is depicted. 
The distributions are not expected to follow Landau-Gauss functions owing to charge sharing and the limitations of the charge measurement and calibration.
The lower seed signal for the pixel flavour with continuous n-implant is a consequence of the higher charge sharing.

The cluster charge is not evaluated quantitatively in this document owing to the limited precision in the conversion of ToT values to physical units as discussed in Section~\ref{sec:configurations}.

\subsection{Hit detection efficiency}
\label{sec:performance_efficiency}

\begin{figure*}[btp]
	\centering
	\begin{minipage}[b]{0.49\textwidth}
		\includegraphics[width=\columnwidth]{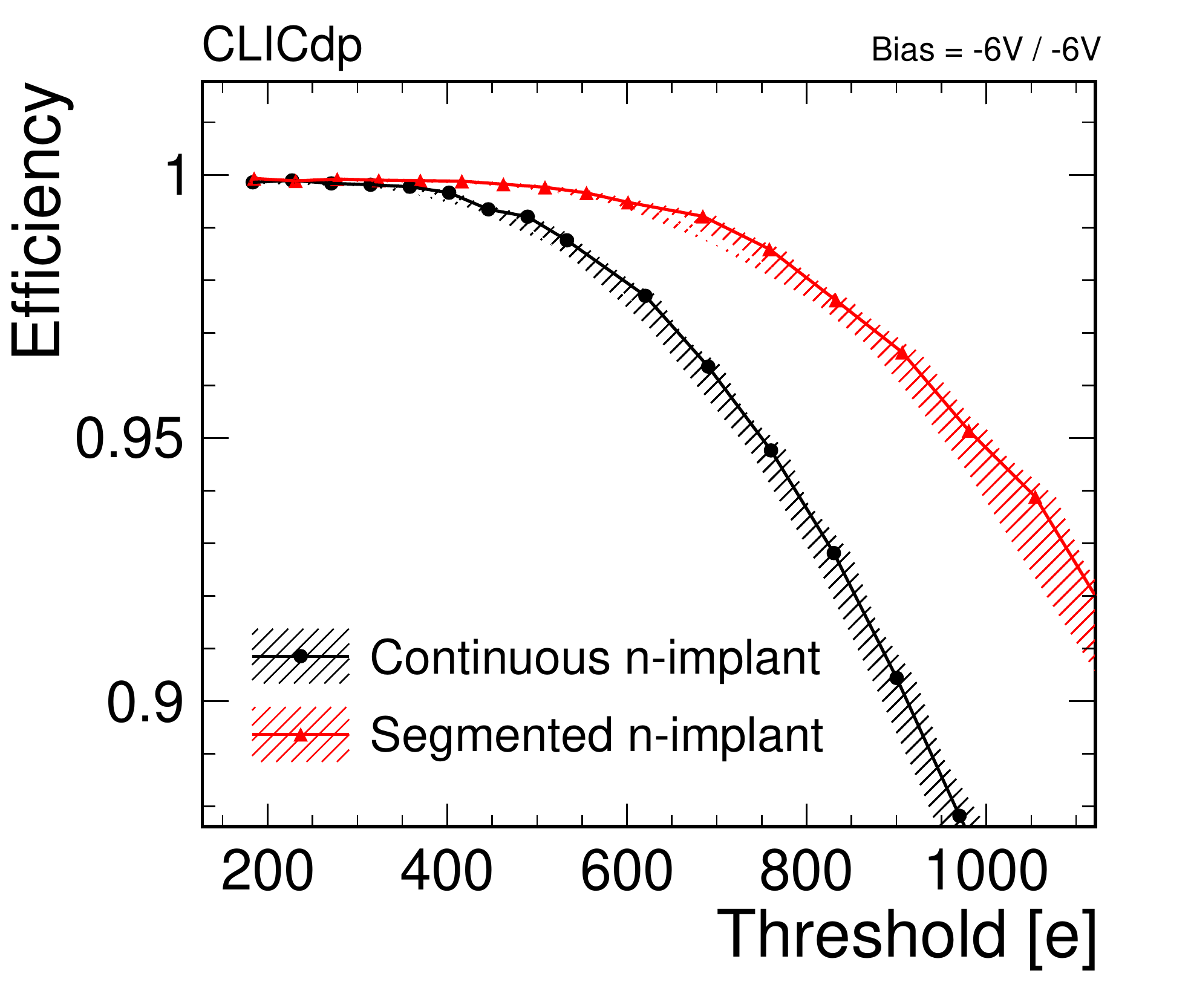}%
		\caption{Hit detection efficiency as a function of threshold for both pixel flavours.
		The hatched band represents the statistical and systematic uncertainties.}
		\label{fig:efficiency_scan}
	\end{minipage}
	\hfill
	\begin{minipage}[b]{0.49\textwidth}
		\includegraphics[width=\linewidth]{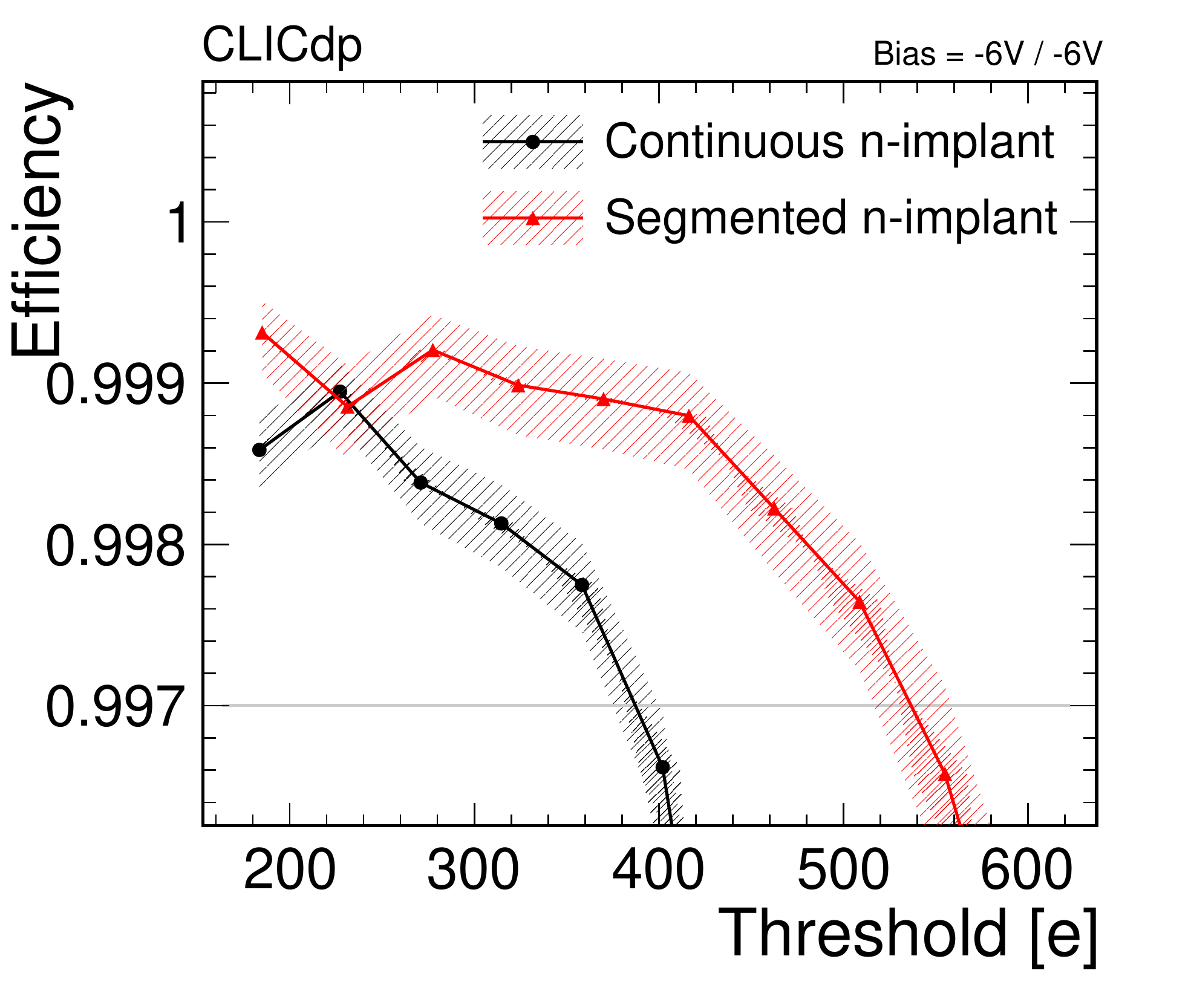}
		\caption{Hit detection efficiency as a function of threshold for low thresholds.
		The hatched band represents the statistical and systematic uncertainties.}
		\label{fig:efficiency_scan_zoomed}
	\end{minipage}
\end{figure*}

The hit detection efficiency as a function of the detection threshold is depicted in Fig~\ref{fig:efficiency_scan}.
A close-up of the high efficiency range is shown in Fig.~\ref{fig:efficiency_scan_zoomed}.
The maximum threshold with an efficiency of $> 99.7\% $ and the range between this value and the operation threshold (defined as \textit{efficient operation window}) are listed in Table~\ref{tab:tb_results}.
For the two thinned assemblies, the operation threshold is 180\,e. 

The statistical uncertainties arise from the uncertainties on the efficiency values.
The systematic uncertainty is given by the uncertainty on the threshold value. 

The efficient operation window evaluates to
$$
207 \pm 12 \textrm{ (stat.)} ^{+ 5}_{- 7}  \textrm{ (syst.)}\,\mathrm{e},
$$
for the continuous n-implant and 
$$
357 \pm 20 \textrm{ (stat.)} ^{+ 8}_{- 11}  \textrm{ (syst.)}\,\mathrm{e}
$$
for the flavour with the segmented n-implant.
The efficient operation window for the second flavour is more than 1.5 times larger owing to the reduced charge sharing, which gives rise to a higher seed signal.

\begin{figure}[tbp]
	\centering
	\includegraphics[width=\columnwidth]{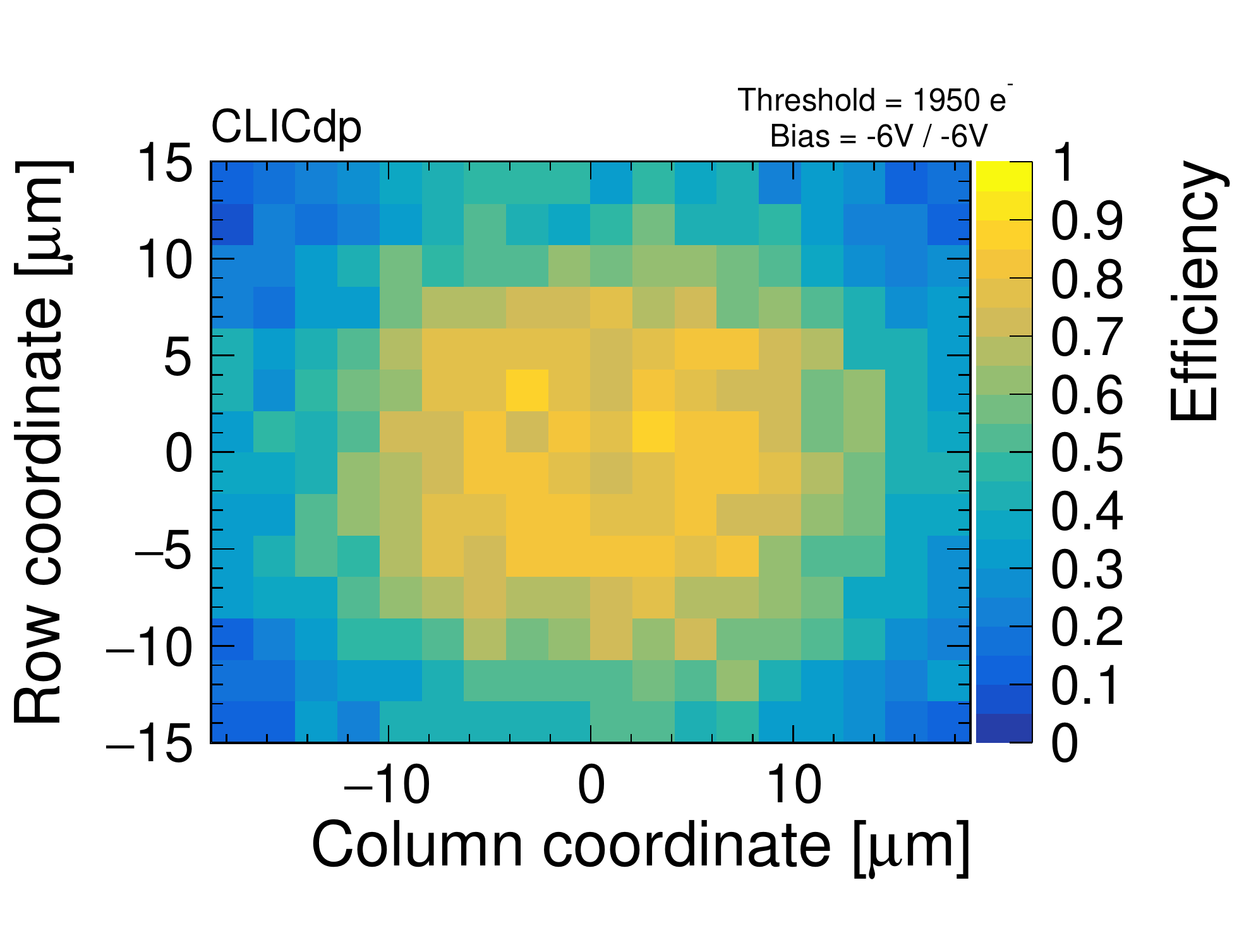}%
	\caption{In-pixel hit detection efficiency at a threshold of 1950\,e$^-$ for the pixel flavour with continuous n-implant.}
	\label{fig:efficiency_450e}
\end{figure}

For high detection thresholds, inefficient regions start to form at the pixel edges, as illustrated in Fig.~\ref{fig:efficiency_450e}, where the in-pixel hit detection efficiency is shown at a threshold of 1950\,e$^-$ for the flavour with continuous n-implant.
The pixel corners are especially affected as a result of enhanced charge sharing in these regions.

\subsection{Spatial resolution}

\begin{figure}[tbp]
	\centering
	\includegraphics[width=\columnwidth]{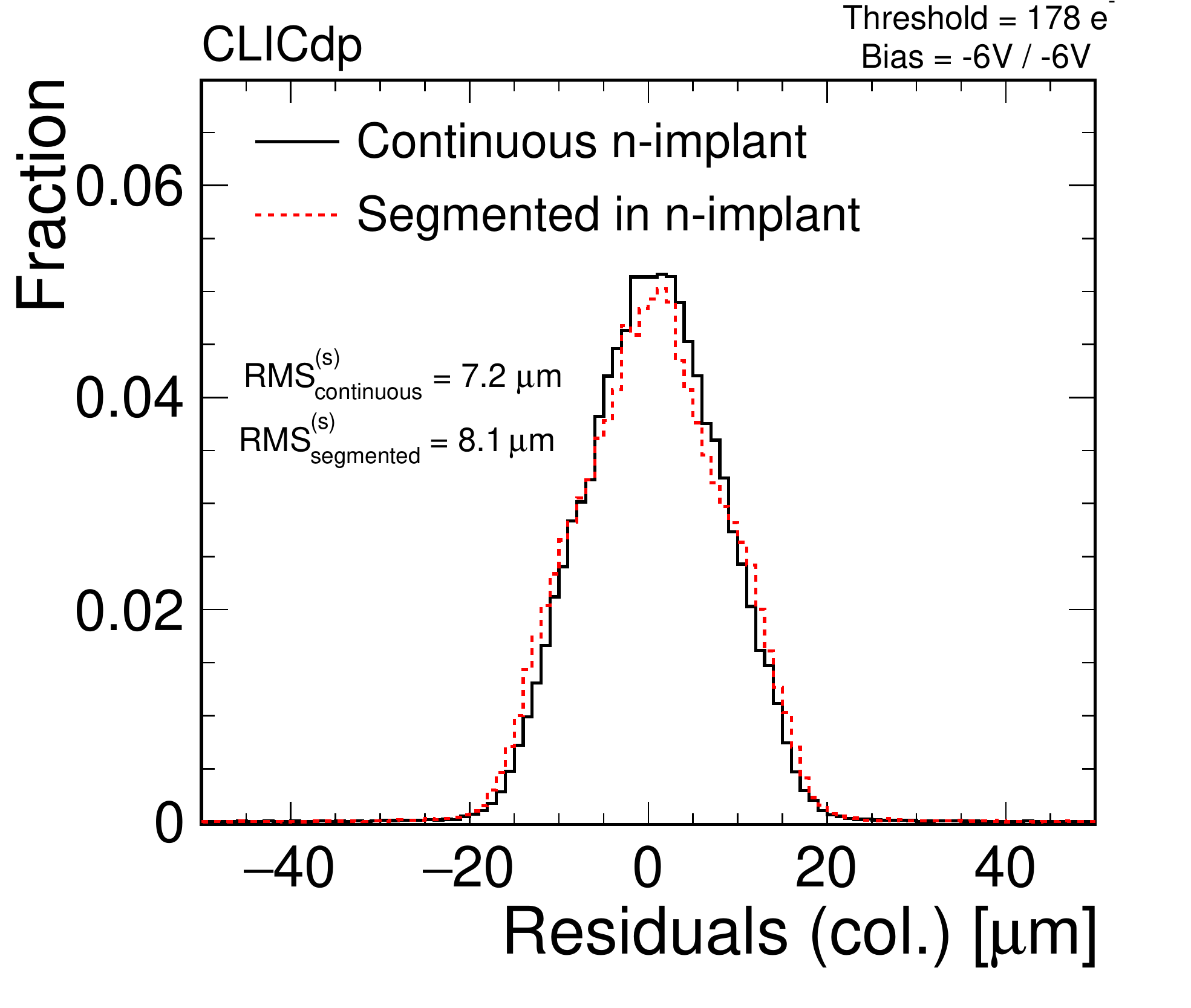}%
	\caption{Residuals in column direction between track intercept position and reconstructed cluster position on the CLICTD.
		The error bars reflecting the statistical uncertainty are not visible.}
	\label{fig:A1_B4_residuals_Y}
\end{figure}

\paragraph{Nominal conditions}
The spatial residuals, defined as the difference between the reconstructed cluster position and the track intercept on the DUT, are shown in Fig.~\ref{fig:A1_B4_residuals_Y} for the column direction.
The RMS of the central 99.7\% of the distribution
amounts to \SI{7.2}{\micro m} for the pixel flavour with continuous n-implant and \SI{8.1}{\micro m} for the one with segmented n-implant.

The spatial telescope track resolution at the DUT is quadratically subtracted from the measured RMS to obtain the spatial resolution of the DUT.

The statistical uncertainty on the spatial resolution is of the order of $1 \times 10^{-2}\,\SI{}{\micro m}$. 
To quantify the systematic uncertainties, the telescope single plane resolution is varied within its uncertainties given in~\cite{Jansen:2016bkd}, which yields an uncertainty of $\pm \SI{0.1}{\micro m}$. 
The  propagated threshold uncertainty is $\pm \SI{0.1}{\micro m}$ as well.
The total systematic uncertainty is given by the quadratic sum of the individual values.

The spatial resolution in row direction evaluates to $4.6 \pm 0.2$\,\SI{}{\micro m} for both pixel flavours.
Observing identical values is in agreement with the similar cluster row size presented in Section~\ref{sec:charge_sharing}.
The value is well below the requirement of $\SI{7}{\micro m}$ for the CLIC tracking detector.

In column direction, the spatial resolution for the pixel flavour with segmented n-implant is $7.6 \pm 0.2$\,\SI{}{\micro m}, which is approximately 12\% larger compared to the $6.7 \pm 0.2$\,\SI{}{\micro m} that is measured for the pixel flavour with continuous n-implant.

In both dimensions, the spatial resolution is superior to the binary resolution that would be expected without charge sharing. 
The binary resolution is given by $\textrm{pitch}/\sqrt{12}$ and evaluates to $\SI{8.7}{ \micro m}$ in row direction and $\SI{10.8}{ \micro m}$ in column direction.

\begin{figure*}[btp]
	\centering
	\begin{minipage}[b]{0.49\textwidth}
		\includegraphics[width=\columnwidth]{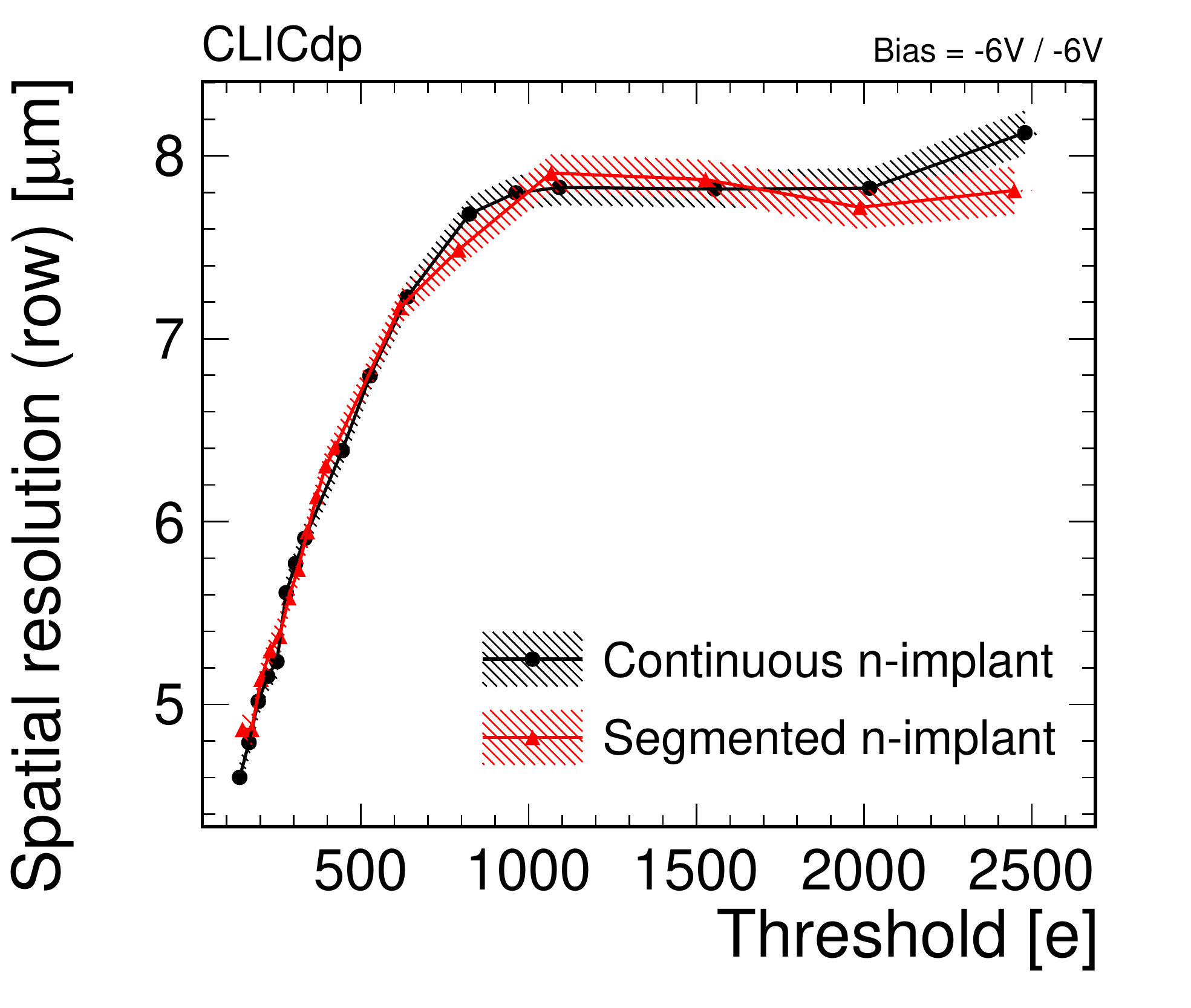}%
		\caption{Spatial resolution in row direction as a function of threshold for both pixel flavours.
			The hatched band represents the statistical and systematic uncertainties.}
		\label{fig:A1_B4_residuals_threshold}
	\end{minipage}
	\hfill
	\begin{minipage}[b]{0.49\textwidth}
		\includegraphics[width=\columnwidth]{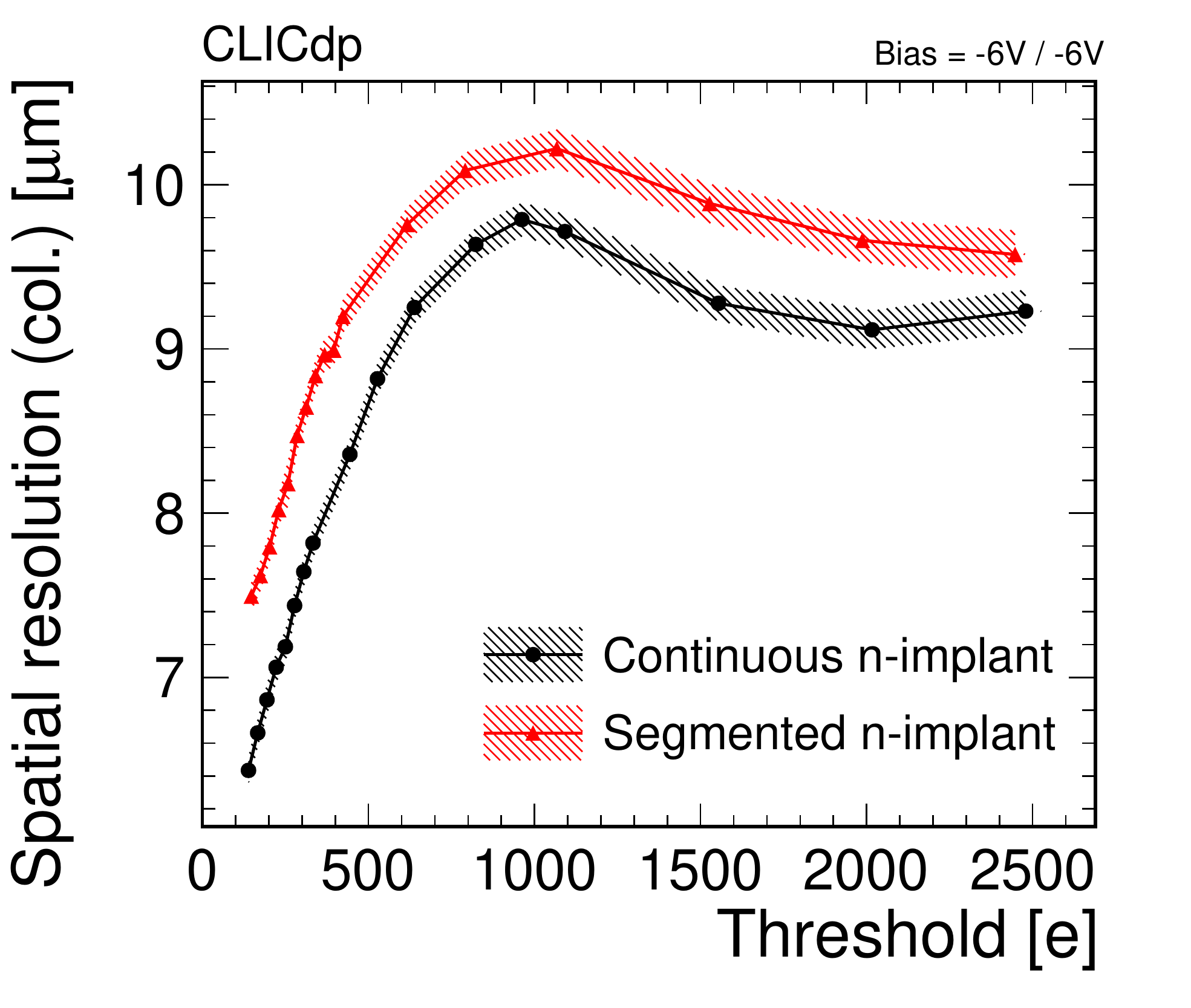}%
		\caption{Spatial resolution in column direction as a function of threshold for both pixel flavours.
			The hatched band represents the statistical and systematic uncertainties.}
		\label{fig:A1_B4_residualsX_threshold}
	\end{minipage}
\end{figure*}

\paragraph{Threshold scan}

In Fig.~\ref{fig:A1_B4_residuals_threshold}, the spatial resolution is shown in row direction as a function of the detection threshold.
With increasing threshold, the spatial resolution degrades owing to the decrease in cluster size.
The binary resolution of $\SI{8.7}{\micro m}$ is never exceeded.

For threshold values greater than 1000\,e$^-$, the efficient pixel area starts to shrink from the pixel edges leading to an improvement in the spatial resolution.

The spatial resolution in column direction as a function of threshold is depicted in Fig.~\ref{fig:A1_B4_residualsX_threshold}.
It illustrates that the reduced charge sharing for the flavour with the segmented n-implant causes a degrading spatial resolution regardless of the detection threshold. 


\subsection{Time resolution}

\begin{figure}[tbp]
	\centering
	\includegraphics[width=\columnwidth]{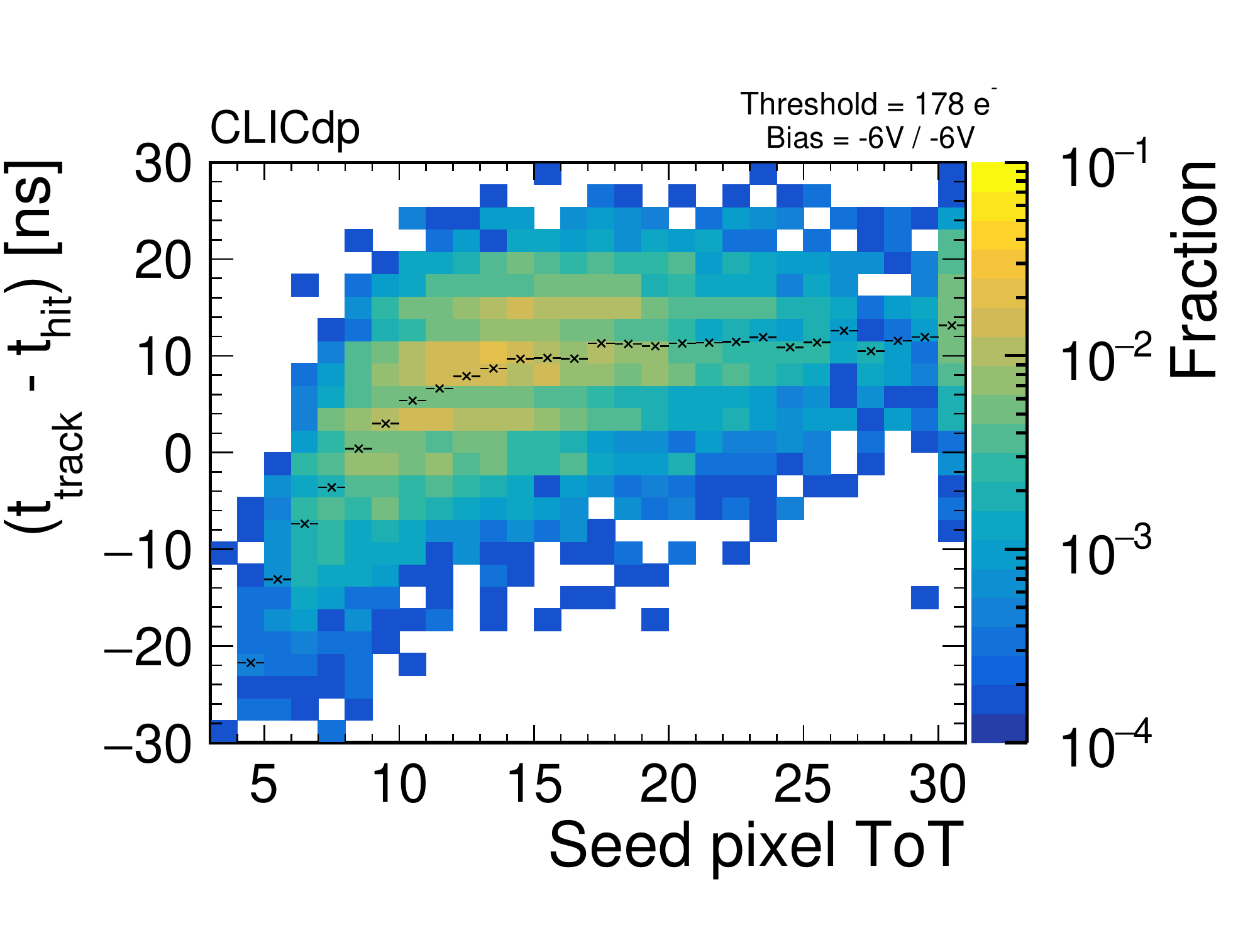}%
	\caption{Residual between track timestamp and reconstructed cluster timestamp as a function of seed pixel charge before time-walk correction for the pixel flavour with continuous n-implant.
	The black crosses denote the mean of each ToT bin.}
	\label{fig:timeWalkCurveBeforeCorrection}
\end{figure}

\begin{figure}[tbp]
	\centering
	\includegraphics[width=\columnwidth]{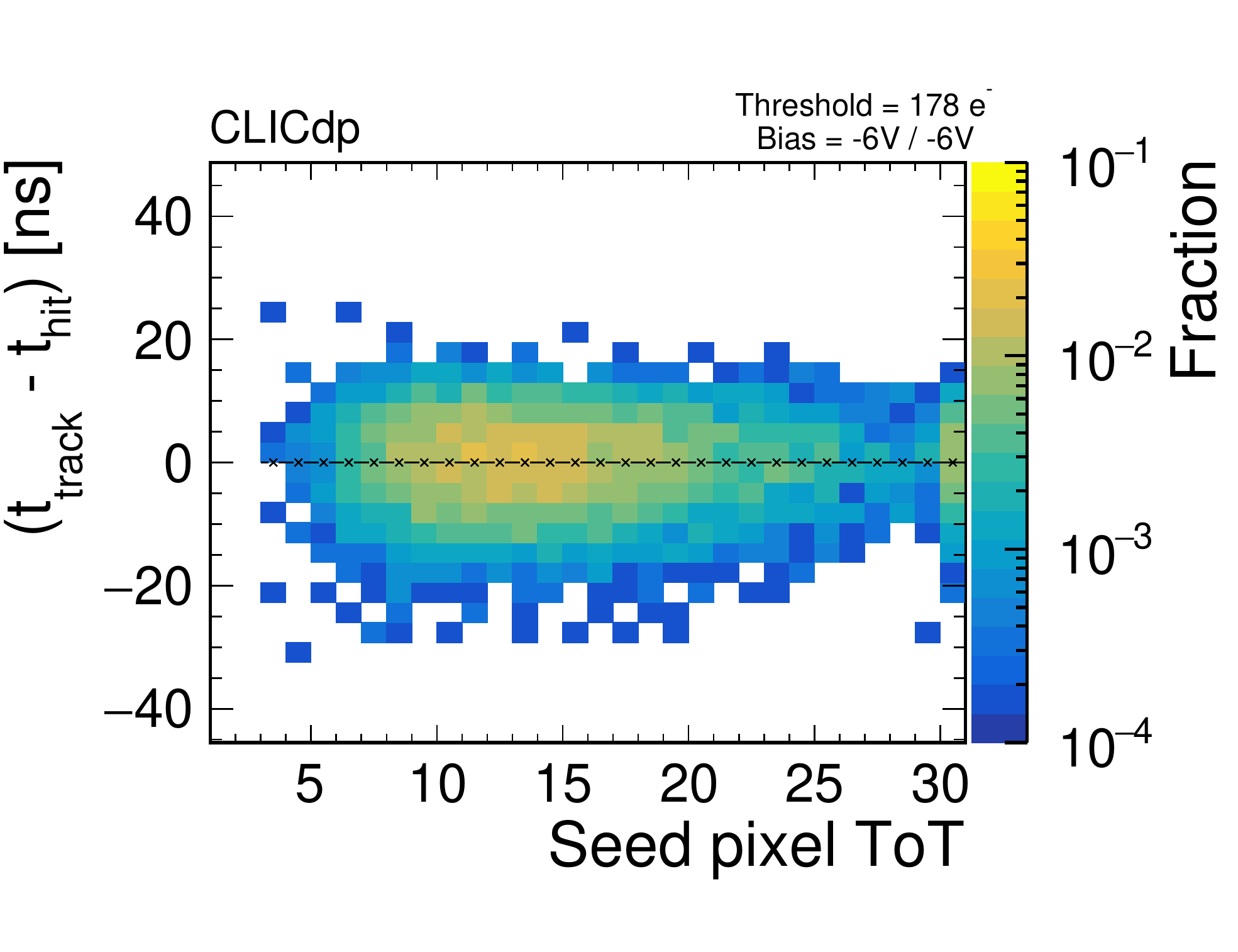}%
	\caption{Residual between track timestamp and reconstructed cluster timestamp as a function of seed pixel charge after time-walk correction for the pixel flavour with continuous n-implant.
	The black crosses denote the mean of each ToT bin.}
	\label{fig:timeWalkCurveAfterCorrection}
\end{figure}

\begin{figure*}[btp]
	\centering
	\begin{minipage}[b]{0.49\textwidth}
		\includegraphics[width=\linewidth]{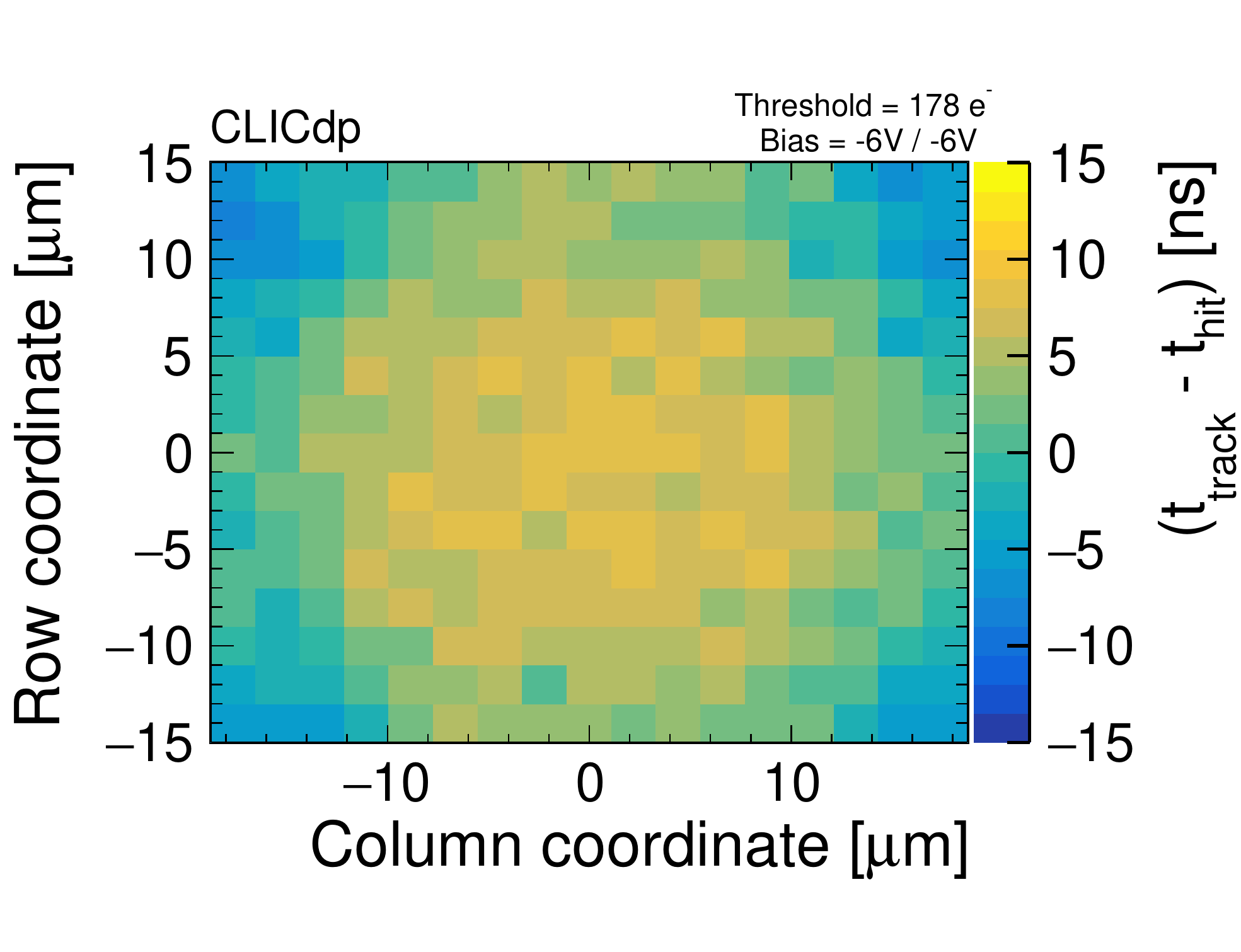}
		\caption{In-pixel time residuals for the pixel flavour with continuous n-implant before time-walk correction.}
		\label{fig:inPixelTimine_A1m6m6}
	\end{minipage}
	\hfill
	\begin{minipage}[b]{0.49\textwidth}
		\includegraphics[width=\linewidth]{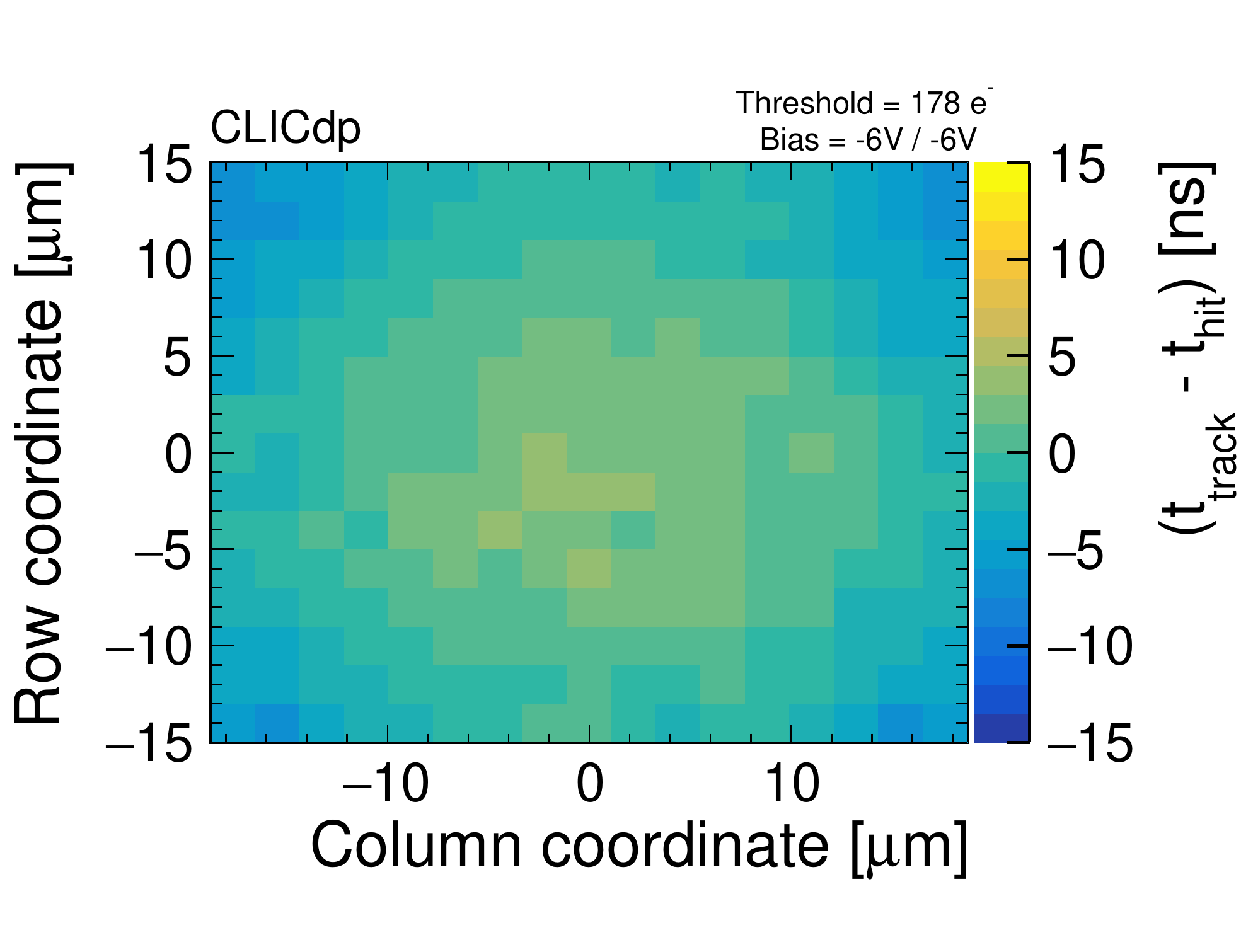}
		\caption{In-pixel time residuals for the pixel flavour with segmented n-implant before time-walk correction.}
		\label{fig:inPixelTimine_B1m6m6}
	\end{minipage}
\end{figure*}

\paragraph{Nominal conditions}
The time residuals are defined as the difference between the track timestamp and the ToA of the DUT.
In Fig.~\ref{fig:timeWalkCurveBeforeCorrection}, the time residuals are depicted as a function of the seed pixel ToT for the pixel flavour with continuous n-implant.
A slower response is observed for low signal heights (\textit{time-walk}).
The effect is particularly strong in the pixel corners, where a lower seed signal is expected, as can be seen in Fig.~\ref{fig:inPixelTimine_A1m6m6} for the continuous n-implant and in Fig.~\ref{fig:inPixelTimine_B1m6m6} for the segmented n-implant.
As a consequence, the time-walk is more pronounced for the flavour with continuous n-implant. 
This result is particularly interesting for applications where precise timing is required before an offline time-walk correction can be applied.

\begin{figure*}[btp]
	\centering
	\begin{minipage}[b]{0.49\textwidth}
		\includegraphics[width=\linewidth]{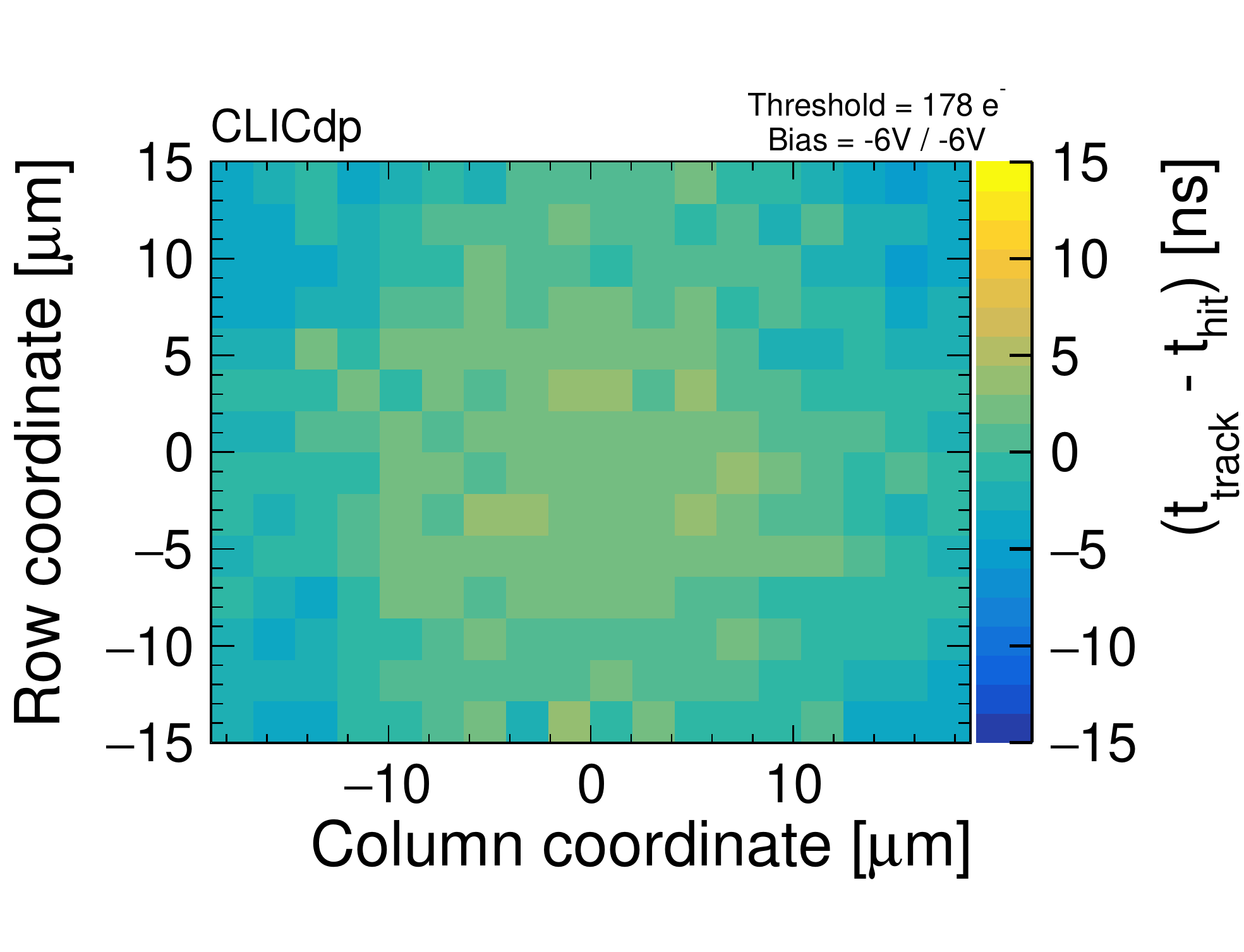}
		\caption{In-pixel time residuals for the pixel flavour with continuous n-implant after time-walk correction.}
		\label{fig:inPixelTimine_A1m6m6_corrected}
	\end{minipage}
	\hfill
	\begin{minipage}[b]{0.49\textwidth}
		\includegraphics[width=\linewidth]{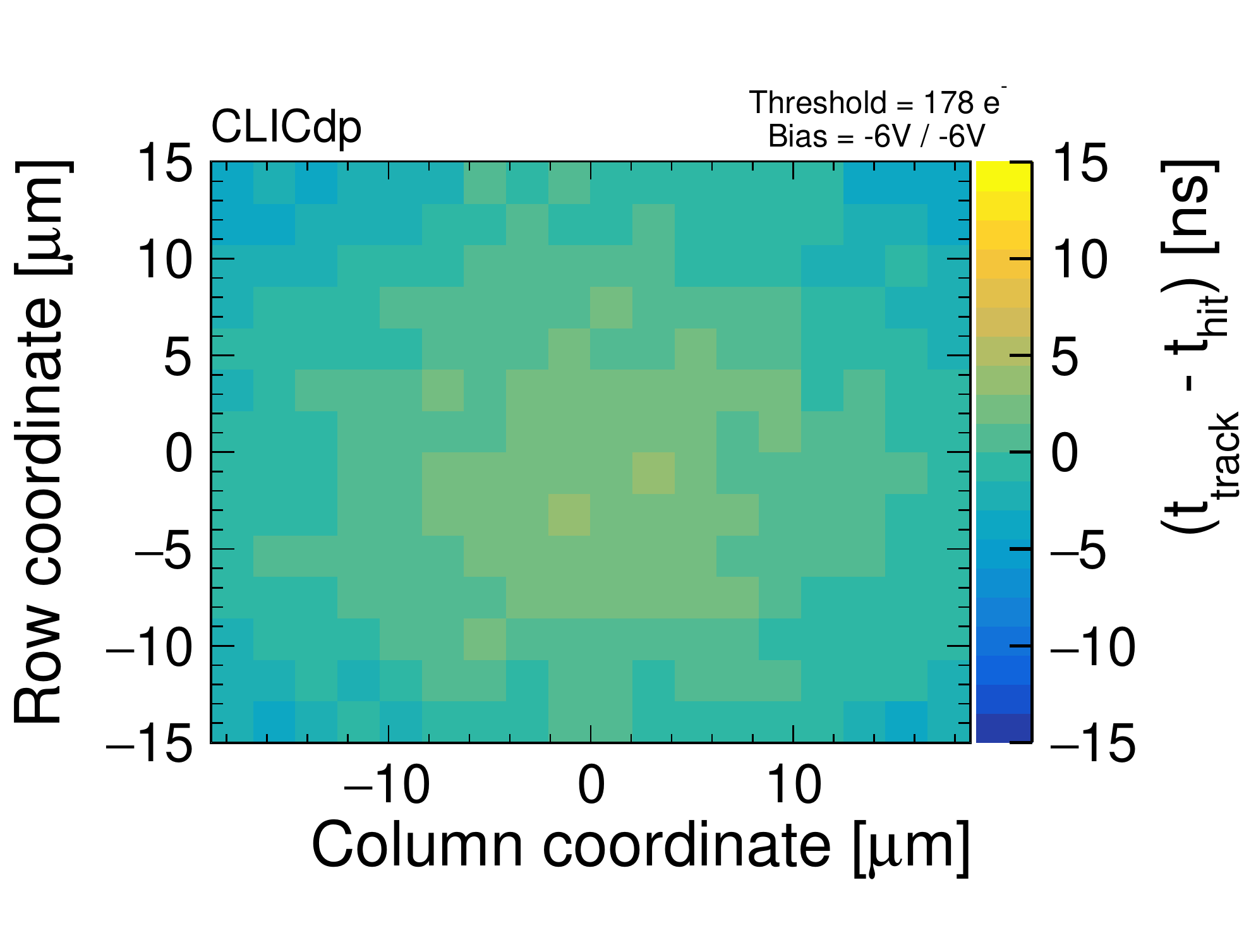}
		\caption{In-pixel time residuals for the pixel flavour with segmented n-implant after time-walk correction.}
		\label{fig:inPixelTimine_B1m6m6_corrected}
	\end{minipage}
\end{figure*}

For the CLICTD sensor, a time-walk correction is performed for each ToT bin separately by subtracting the mean time difference between the track and the measured ToA. 
The time residuals after correction as a function of the seed pixel ToT are shown in Fig.~\ref{fig:timeWalkCurveAfterCorrection}.
The width of the time residuals is larger for small seed signals due to the stronger impact of amplitude jitter.
The in-pixel time residuals after time-walk correction are depicted in Figs.~\ref{fig:inPixelTimine_A1m6m6_corrected} and~\ref{fig:inPixelTimine_B1m6m6_corrected} for continuous n-implant and segmented n-implant, respectively. 
After time-walk correction, the timing is more homogenous across the pixel cell for both pixel flavours.
The remaining in-pixel pattern suggests that slow signals arise predominantly from incident positions at the pixel corners, which is in agreement with 3D TCAD simulations indicating that the time resolution degrades in the pixel edge regions~\cite{Munker_2019}. 

\begin{figure}[tbp]
	\centering
	\includegraphics[width=\columnwidth]{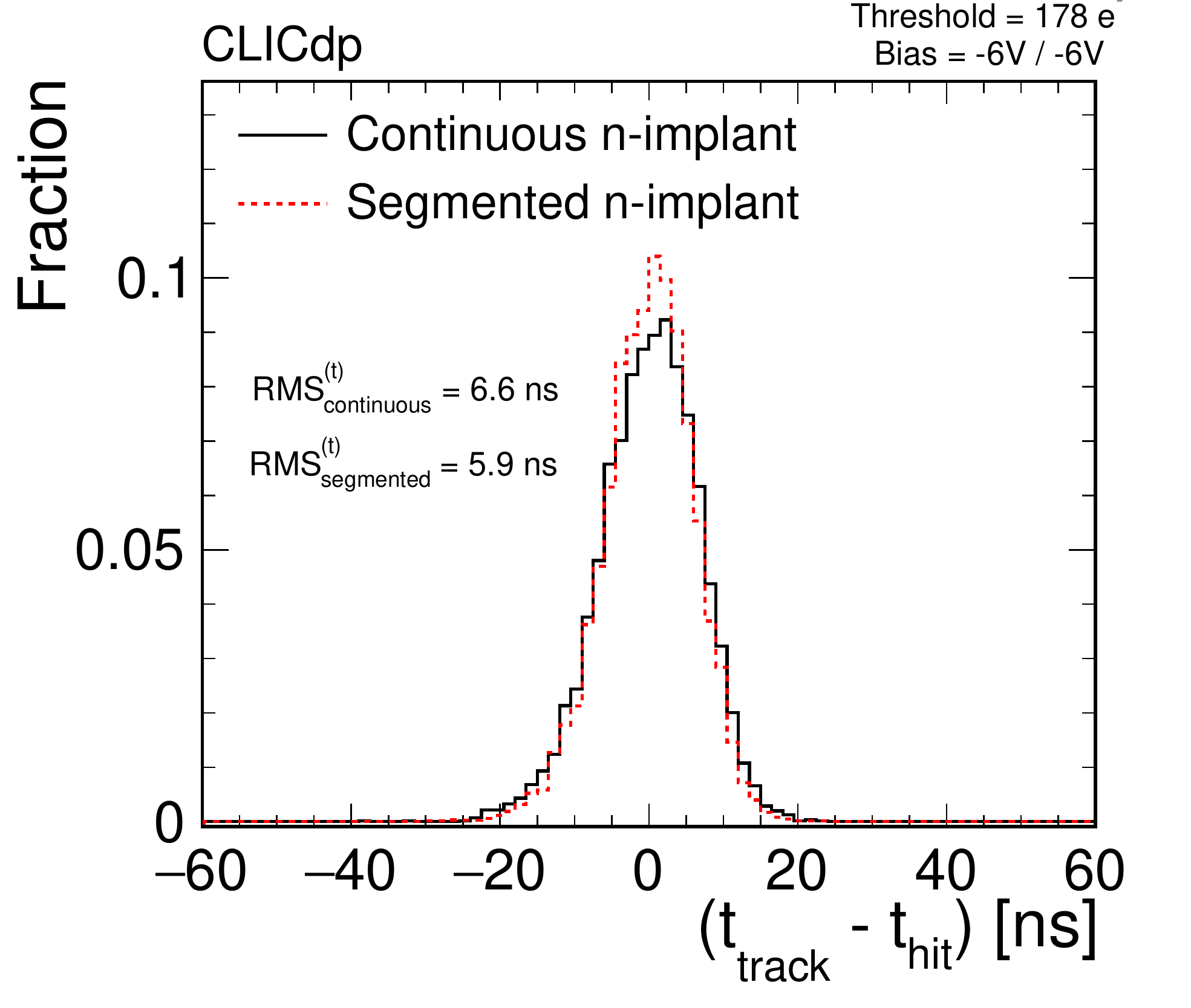}%
	\caption{Time residuals between track timestamp and CLICTD timestamp after time-walk correction.
    The error bars reflecting the statistical uncertainty are not visible.}
	\label{fig:A1_B1_timing}
\end{figure}

The one dimensional residual distributions are depicted in Fig.~\ref{fig:A1_B1_timing} for both pixel flavours.  
The time resolution is calculated using the RMS of the central 99.7\% of the distribution, which amounts to 6.6\,ns and 5.9\,ns for the pixel flavour with continuous n-implant and segmented n-implant, respectively. 

The time resolution associated with the track timestamp ($1.1$\,ns) is quadratically subtracted from the RMS.

The statistical uncertainties as well as the propagated threshold uncertainty are of the order of $\SI{0.01}{ns}$. 
The systematic uncertainty is estimated by repeating the time-walk correction for every sub-pixel position in a channel individually.
The spread of the sub-pixel specific time resolution is $\pm 0.1$\,ns.

The time resolution for the pixel flavour with the segmented n-implant evaluates to $(5.8 \pm 0.1)$\,ns, which is about 10\% better than the $(6.5 \pm 0.1)$\,ns for the flavour with continuous n-implant. 
The values are within the requirements for the CLIC tracking detector. 
The similar timing performance for both pixel flavours, despite the accelerated charge collection for the flavour with segmented n-implant, can be attributed to front-end timing limitations as explained in Section~\ref{sec:configurations}.

\begin{figure}[tbp]
	\centering
	\includegraphics[width=\columnwidth]{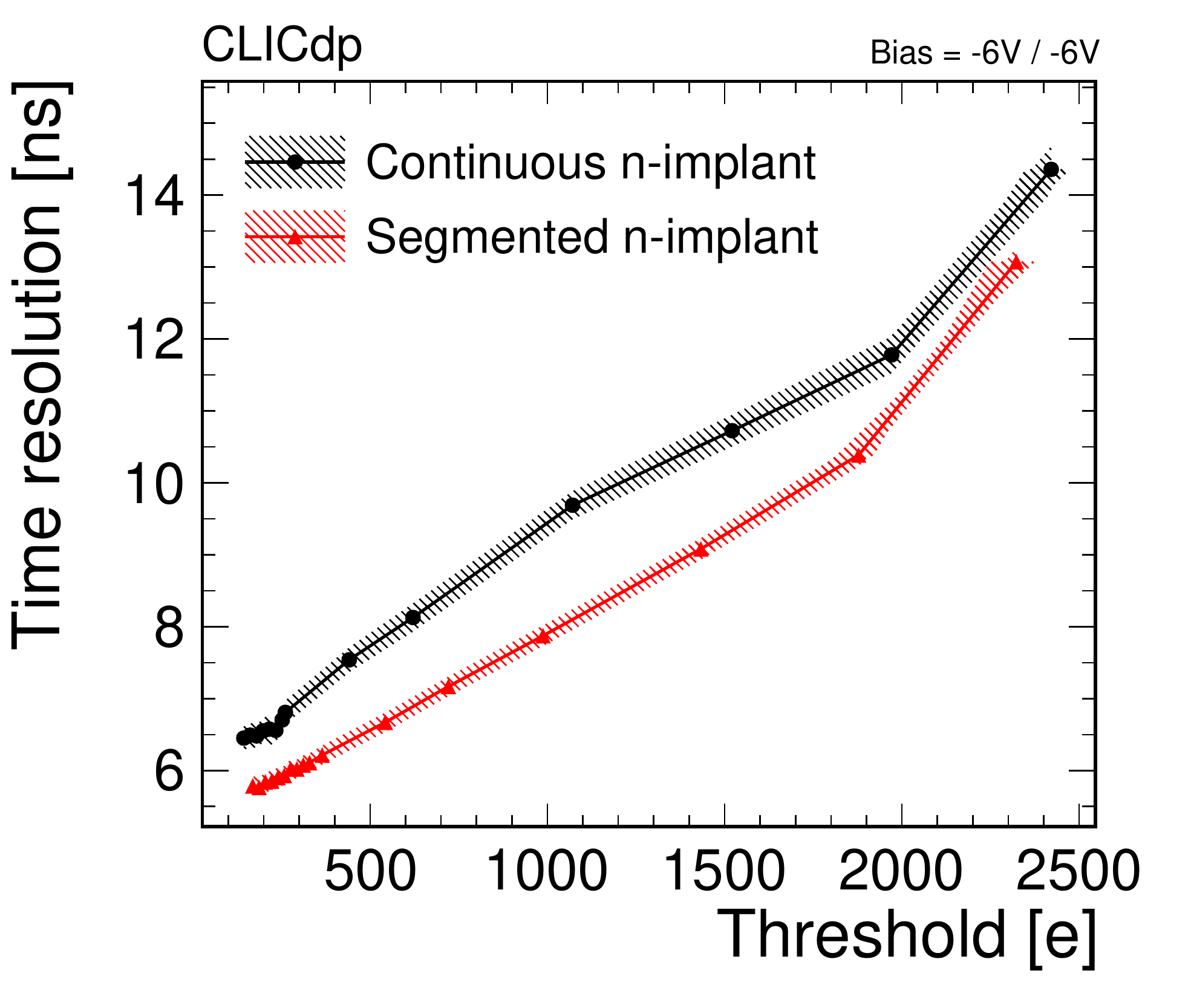}%
	\caption{Time resolution as a function of the detection threshold.
	The hatched band represents the statistical and systematic uncertainties.}
	\label{fig:timing_resolution_scan}
\end{figure}

\paragraph{Threshold scan}
In Fig.~\ref{fig:timing_resolution_scan}, the time resolution is depicted as a function of the detection threshold.
For both pixel flavours, the time resolution increases with increasing threshold  owing to the stronger contribution of amplitude jitter.

%% file: summary.tex
The CLICTD monolithic pixel sensor has been characterised in a charged particle beam for two different pixel flavours. 
In one flavour, a deep continuous low-dose n-implant ensures full lateral depletion of the epitaxial layer. 
In a second flavour, the n-implant is segmented to enhance the lateral electric field for accelerated charge collection and reduced charge sharing.

The requirements for the CLIC tracker in terms of spatial and timing resolution as well as hit detection efficiency are fulfilled.
In addition, the sub-pixel segmentation scheme of the front-end has shown to deliver the required accuracy while minimizing the digital footprint. 
Previous results have shown that the estimated power-consumption~\cite{clictd_design_characterization} as well as the material budget\cite{dort2020clictd} also comply with the requirements.

The measurements confirm that charge sharing is affected by the pixel flavour, but only, as intended, in the column dimension where the segmentation was introduced. 
The position resolution of $\SI{4.6}{\micro m}$ in the other direction remains unaffected. 
The reduced charge sharing for the pixel flavour with the segmented n-implant improves the measured time resolution by \SI{10}{\percent} to $\SI{5.8}{ns}$, despite the limitations in the front-end time resolution. 
Due to the larger seed signal, it also increases the operation window for efficient detection by approximately $\SI{60}{\percent}$, which is particularly important for future sensors with thin active layers.

%% file: credit_statement.tex
\textbf{R. Ballabriga} Methodology, Supervision
\textbf{E. Buschmann} Investigation, Software
\textbf{M.~Campbell} Methodology
\textbf{D.~Dannheim} Investigation, Methodology, Supervision, Writing - Review \& Editing
\textbf{K.~Dort} Formal analysis, Investigation, Software, Visualization, Writing - Original Draft
\textbf{N. Egidos} Resources
\textbf{L. Huth} Investigation
\textbf{I.~Kremastiotis} Investigation, Resources
\textbf{J.~Kr\"oger} Investigation, Software
\textbf{L.~Linssen} Project administration, Funding acquisition
\textbf{X.~Llopart} Resources
\textbf{M.~Munker} Investigation, Methodology, Supervision, Writing - Review \& Editing
\textbf{A.~N\"urnberg} Resources
\textbf{W.~Snoeys} Conceptualization, Resources
\textbf{S.~Spannagel} Investigation, Methodology, Software
\textbf{T.~Vanat} Resources, Software
\textbf{M.~Vicente} Investigation
\textbf{M.~Williams} Investigation, Software

%% file: acknowledgements.tex
This work has been sponsored by the Wolfgang Gentner Programme of the German Federal Ministry of Education and Research (grant no. 05E15CHA).
The measurements leading to these results have been performed at the Test Beam Facility at DESY Hamburg (Germany), a member of the Helmholtz Association (HGF).
This project has received funding from the European Union’s Horizon 2020 research and innovation programme under grant agreement No 654168.
This work was carried out in the framework of the CLICdp Collaboration.